\documentclass[preprint,superscriptaddress,amsmath,amssymb,aps,prd,longbibliography,floatfix,nofootinbib,11pt]{revtex4-2}
\usepackage{graphicx}   
\usepackage[
colorlinks=true,        
citecolor=blue,         
linkcolor=blue,         
urlcolor=blue          
]{hyperref}             
\usepackage{bm}        
\usepackage{xcolor}    
\usepackage{multirow}
\usepackage{mathrsfs}
\usepackage{amsmath}

\begin{document}

\title{Rational Orbits and Gravitational Waves in Static Spherical Spacetimes: An Open-Source Numerical Framework}

\author{Dan Li}
\affiliation{School of Mathematics and Physics, University of South China, Hengyang, 421001 People's Republic of China}

\author{Shiyang Hu}
\email{husy\_arcturus@163.com}
\affiliation{School of Mathematics and Physics, University of South China, Hengyang, 421001 People's Republic of China}

\author{Chen Deng}
\affiliation{School of Astronomy and Space Science, Nanjing University, Nanjing, 210023 People's Republic of China}

\author{Shijie Tan}
\affiliation{School of Mathematics and Physics, University of South China, Hengyang, 421001 People's Republic of China}

\author{Guansheng He}
\affiliation{School of Mathematics and Physics, University of South China, Hengyang, 421001 People's Republic of China}

\begin{abstract}
Timelike orbits constitute a crucial probe for exploring the intrinsic properties of curved spacetimes, and the carried gravitational radiation signals provide a direct window into strong field gravity. In this paper, we develop a versatile computational framework based on Mathematica and the OpenMP parallel architecture to simulate the rational orbits of timelike particles and their gravitational radiation in static spherically symmetric spacetimes. Specifically, requiring only the user defined covariant metric, this numerical tool can efficiently calculate rational orbits across various configurations, as well as the corresponding gravitational wave polarization states and characteristic strains. The package presented here offers a highly efficient and comprehensive one-stop solution for investigating the properties of curved spacetimes and their potential observational signatures. To demonstrate the reliability and capability of our code, we apply it to the Schwarzschild spacetime as a test case, illustrating the functionality of the code across several key aspects, including the effective potential, stable orbital regions, rational and irrational orbits, and gravitational wave signals. Furthermore, we show that the gravitational waves emitted by an extreme-mass-ratio inspiral system composed of an intermediate mass black hole and the Galactic Center supermassive black hole have the potential to be identified by future space detectors.
\end{abstract}

\maketitle

\section{Introduction}
From the detection of gravitational wave signals emitted by binary black hole mergers by the Laser Interferometer Gravitational-Wave Observatory (LIGO) \cite{2016PhRvL.116f1102A}, to the successive imaging of the supermassive black holes in the M87 galaxy and at the Galactic Center (Sgr A$^{*}$) by the Event Horizon Telescope (EHT) \cite{2019ApJ...875L...1E,2022ApJ...930L..12E}, and further to the observational testing of Hawking's area law \cite{2021PhRvL.127a1103I}, the continuously accumulating evidence has made the existence of black holes practically indisputable. Consequently, serving both as natural sources of extreme gravitational fields and as crucial building blocks for refining the theory of gravity, black holes have attracted extensive attention from the scientific community.

Owing to the intense gravitational fields of black holes, small celestial bodies can be gravitationally bound to orbit around them. These orbital motions exhibit a rich variety of patterns, including standard Keplerian orbits, precessing orbits, chaotic motions \cite{2009ApJ...693..472T,2016EPJC...76...32S,2019EPJC...79..479P,2019EPJP..134...96L,2021ApJS..257...40H,2021EPJC...81..785S,2022ApJ...925..158H,2023EPJC...83..828Z,2025EPJC...85..770X}, off-equatorial motions \cite{2014ApJ...787..117K}, and extremely accelerated motions in polar regions \cite{2010PhRvL.104b1101J,2010PhRvD..82j3005W,2016PhRvD..93h4025G}. Regardless of the specific dynamic behavior, the characteristics of these orbits inherently encode the fundamental properties of the background spacetime, making them a premier probe for investigating black holes. Interestingly, by studying the geodesic motion of timelike particles in black hole spacetimes, Levin and Perez-Giz established a numerical framework for calculating strictly closed timelike orbits \cite{2008PhRvD..77j3005L}. Within a complete orbital cycle, the azimuthal and radial periods of such orbits exhibit an integer ratio, hence they are referred to as rational orbits. Specifically, the configuration of a rational orbit can be strictly characterized by three integers $(z,w,v)$, which makes it possible to extract richly structured orbital information within curved spacetimes. Subsequently, this methodology has been widely applied by researchers to various modified theories of gravity \cite{Alloqulov:2025ucf,Misra:2010pu,Wei:2019zdf,Liu:2018vea,Deng:2020yfm,Deng:2020hxw,Lin:2021noq,Wang:2022tfo,Gao:2020wjz,Lin:2022wda,Zhang:2022zox,Gao:2021arw,Lin:2022llz,Lin:2023rmo,QiQi:2024dwc,Lim:2024mkb,Bragado:2025jrg,Shaymatov:2023jfa,Xamidov:2025oqx,Tu:2023xab,Meng:2024cnq,Li:2024tld,Wang:2025hla,Lu:2025xlp,Lu:2025cxx,Zahra:2025tdo,Ahmed:2025azu,Deng:2025wzz,Alloqulov:2025dqi,Yang:2024lmj,Alloqulov:2025bxh,Zhao:2024exh,Shabbir:2025kqh,Haroon:2025rzx,Zhang:2025wni,Junior:2024tmi,Zare:2025aek,Huang:2025vpi,Shabbir:2026qlh,Hua:2026kvw,Li:2025sfe,Huang:2024oli,Huang:2025czx,Jiang:2024cpe,Huang:2026oga,Kumar:2026hfx}, establishing a direct connection between spacetime parameters and orbital configurations. This ultimately provides an extensive theoretical catalog for observing timelike orbits, refining theories of gravity, and constraining spacetime parameters.

It is worth noting that rational orbits are not merely mathematical deductions; they not only encode spacetime information but also play a significant role in orbital resonance and nonlinear theories. Therefore, further exploring the potential observational signatures of rational orbits holds substantial practical importance. Interestingly, a small celestial body moving on a rational orbit and the central black hole can constitute an extreme-mass-ratio inspiral system, whose mass quadrupole moment evolves with time, thereby generating gravitational radiation. Due to the breakdown of the adiabatic evolution condition at resonances, these radiation signals carry unique fingerprint features, providing a crucial yardstick for inversely extracting the intrinsic properties of spacetime. More importantly, the gravitational wave signals of such systems typically fall within the mHz frequency window, serving as a core scientific target for future space gravitational wave observatories, such as LISA \cite{Robson:2018ifk,Caprini:2025mfr}, TianQin \cite{2016CQGra..33c5010L,2019PhRvD.100d4042L,Li:2024rnk}, and Taiji \cite{Ruan:2018tsw,Liu:2026bbp}. This undoubtedly further solidifies the necessity of establishing a library of gravitational wave signals from rational orbits, utilizing them as a unique probe for testing gravity in the strong field regime.

Despite the extensive and enthusiastic research devoted to rational orbits in curved spacetimes, a universal and comprehensive open-source numerical tool capable of efficiently and conveniently simulating these orbits and their associated gravitational wave signals has remained absent to date. This deficiency implicitly impedes further detailed exploration within the field. To address this gap, we have developed a computational platform based on Mathematica and the OpenMP computing framework to realize the aforementioned functionalities, making it seamlessly applicable to any static spherically symmetric spacetime. Specifically, requiring the user to define only the covariant metric, the code directly returns circular orbits, rational orbits, irrational orbits, and the specific energy range corresponding to the stable orbital regions, alongside the gravitational wave signals and characteristic strains under the designated parameters.

The remainder of this paper is organized as follows. In the next section, we detail the programming logic and fundamental formulas of the open-source code, mathematically demonstrating how to derive rational orbits, the corresponding gravitational wave signals, and characteristic strains starting from a static spherically symmetric spacetime line element. In the section III, we adopt the Schwarzschild spacetime as a test case to showcase the versatile functionalities of the code, focusing on the observational prospects of gravitational wave signals emitted by an extreme-mass-ratio inspiral system composed of the Galactic Center black hole and a small celestial body. In the final section, we draw our conclusions and provide a discussion. Throughout the mathematical derivations in this paper, we adopt geometrized units, where the black hole mass $M$, the gravitational constant $G$, and the speed of light $c$ are all set to unity ($M=G=c=1$). When considering practical applications, we restore the relevant quantities to the International System of Units.
\section{Mathematical Foundation of the Code}
\subsection{Spacetime Model and Equations of Motion for Timelike Particles}
We consider a static spherically symmetric spacetime, whose line element in the coordinates $x^{\mu}=(t,r,\theta,\varphi)$ is written as
\begin{equation}\label{1}
\textrm{d}s^{2} = g_{tt}\textrm{d}t^{2} + g_{rr}\textrm{d}r^{2} + g_{\theta\theta}\textrm{d}\theta^{2} + g_{\varphi\varphi}\textrm{d}\varphi^{2},
\end{equation}
where $g_{\mu\nu}$ the covariant metric, with $f(r) = -g_{tt}$ representing the metric potential. We introduce the four-velocity $\dot{x}^{\mu}=(\dot{t},\dot{r},\dot{\theta},\dot{\varphi})$. Then, the Lagrangian $\mathcal{L}$ governing the particle motion in the spacetime is written as
\begin{eqnarray}\label{2}
\mathcal{L} = \frac{1}{2}g_{\mu\nu}\dot{x}^{\mu}\dot{x}^{\nu} = \frac{1}{2}\left(g_{tt}\dot{t}^{2}+g_{rr}\dot{r}^{2}+g_{\theta\theta}\dot{\theta}^{2}+g_{\varphi\varphi}\dot{\varphi}^{2}\right).
\end{eqnarray}

Since the spacetime is static and spherically symmetric, the line element \eqref{1} does not explicitly depend on the time $t$ and the azimuthal angle $\varphi$, which leads to two conserved quantities for the particle motion, namely the specific energy $E$ and the specific angular momentum $L$. According to the Euler-Lagrange equations, the mapping between these conserved quantities and the velocities is given by
\begin{eqnarray}
-E=p_{t}=g_{tt}\dot{t}, \label{3} \\
L=p_{\varphi}=g_{\varphi\varphi}\dot{\varphi}. \label{4}
\end{eqnarray}
Here, $p_{\mu}$ represents the conjugate momentum corresponding to the generalized coordinates of the particle. Subsequently, we have
\begin{eqnarray}
p_{r}=g_{rr}\dot{r}, \label{5} \\
p_{\theta}=g_{\theta\theta}\dot{\theta}. \label{6}
\end{eqnarray}

Based on the Legendre transformation, we can obtain the Hamiltonian governing the particle motion in the spacetime as
\begin{eqnarray}\label{7}
\mathcal{H} = p_{\mu}\dot{x}^{\mu}-\mathcal{L} = \frac{1}{2}g^{\mu\nu}p_{\mu}p_{\nu},
\end{eqnarray}
where $g^{\mu\nu}$ denotes the contravariant metric, and in the spherically symmetric case, $g^{\mu\nu}=1/g_{\mu\nu}$. Given the initial conditions of the timelike particle $(t,r,\theta,\varphi)$ and $(p_{t},p_{r},p_{\theta},p_{\varphi})$, the particle orbits can be obtained according to the canonical equations
\begin{eqnarray}\label{8}
\dot{x^{\mu}} = \frac{\partial\mathcal{H}}{\partial p_{\mu}}, \quad \dot{p_{\mu}} = - \frac{\partial\mathcal{H}}{\partial x^{\mu}}.
\end{eqnarray}
\subsection{Circular Orbits and Precessing Orbits}
We discuss the orbital solutions of timelike particles in the equatorial plane, so we can safely set $\theta=\pi/2$ and $\dot{\theta}=p_{\theta}=0$, along with the additional conserved quantity of the particle $\mathcal{L}=-1/2$. Substituting these conditions into the Lagrangian, we obtain
\begin{equation}\label{9}
\dot{r}^{2}+V_{\textrm{eff}}^{2}=\frac{-E^{2}}{g_{tt}g_{rr}},
\end{equation}
where $V_{\textrm{eff}}$ represents the particle effective potential\footnote{For certain spacetimes, the effective potential might exhibit multiple potential wells and barriers. It should be emphasized that our current code is exclusively applicable to spacetimes where the effective potential contains at most a single potential well (specifically, the effective potential possesses at most one local maximum and one local minimum). The extension accommodating spacetimes with multiple potential wells is planned to be implemented in future updates of the code.}, written as
\begin{equation}\label{10}
V_{\textrm{eff}}=\sqrt{\frac{1}{g_{rr}}\left(1+\frac{L^{2}}{g_{\varphi\varphi}}\right)}.
\end{equation}

For standard circular orbits in the spacetime, the condition $\dot{r}=0$ must be satisfied. Specifically, there exist two types of special circular orbits in the spacetime, namely the innermost stable circular orbit (ISCO) and the marginally bound circular orbit (MBO). The former corresponds to the minimum point of the effective potential and can be determined by solving
\begin{eqnarray}
\frac{\partial V_{\textrm{eff}}}{\partial r}=0, \label{11} \\
\frac{\partial^{2} V_{\textrm{eff}}}{\partial r^{2}}=0, \label{12}
\end{eqnarray}
to obtain the orbital radius $r_{\textrm{isco}}$ and the corresponding orbital angular momentum $L_{\textrm{isco}}$. Furthermore, by substituting the roots of equations \eqref{11} and \eqref{12} into equation \eqref{9}, we can obtain the specific energy $E_{\textrm{isco}}$ corresponding to the ISCO. It is worth pointing out that for metrics of the type $g_{tt}=-1/g_{rr}$, the $V_{\textrm{eff}}$ corresponding to $r_{\textrm{isco}}$ is exactly the specific energy, which is clear in the radial evolution equation \eqref{9}.

As for the MBO, it corresponds to the maximum point of the effective potential and is therefore unstable; any perturbation can cause a timelike particle on the MBO to plunge into the black hole or escape to infinity. Based on condition \eqref{11} and \eqref{9}, with $E_{\textrm{mbo}}=1$, we can obtain the orbital radius $r_{\textrm{mbo}}$ and $L_{\textrm{mbo}}$ of the MBO.

For any arbitrary angular momentum $L_{i}$ belonging to the interval $(L_{\textrm{isco}},L_{\textrm{mbo}})$, the effective potential \eqref{10} possesses both a maximum point and a minimum point, corresponding to the unstable circular orbit (UnSCO) and the stable circular orbit (SCO) for $L_{i}$. Their respective radii can be obtained by individually solving equation \eqref{11} under the premise of a given $L_{i}$. Subsequently, by substituting the two radii and $L_{i}$ into equation \eqref{9} respectively, we can acquire the specific energies corresponding to the UnSCO and the SCO, denoted as $E_{\textrm{max}}$ and $E_{\textrm{min}}$ respectively. More importantly, by arbitrarily selecting $E_{i}$ within the interval $(E_{\textrm{min}},E_{\textrm{max}})$, the combination of $E_{i}$ and $L_{i}$ corresponds to a stable precessing orbit. In other words, substituting $E_{i}$ and $L_{i}$ into the equation $\dot{r}^{2}=0$ yields three real roots. Arranged in descending order as $r_{2}$, $r_{1}$, and $r_{0}$, the orbit is confined to oscillate within the range $[r_{1},r_{2}]$, where $r_{1}$ and $r_{2}$ correspond to the periastron and apastron of the orbit respectively, while $r_{0}$ lacks physical significance and is merely mathematically valid.
\subsection{Rational Orbits}
It is not difficult to find that the combinations of $E_{i}$ and $L_{i}$ are infinite, and in most cases, the corresponding precessing orbits are not closed. That is to say, after the timelike particle completes one period in the azimuthal direction, its radial coordinate cannot return to the initial value. As the orbital evolution time extends, the envelope of the precessing orbit eventually forms a circle. Interestingly, within these infinite combinations, there exist certain special cases where closed precessing orbits emerge, which are referred to as rational orbits. Specifically, the period $T_{\varphi}$ of the azimuthal angle $\varphi$ of the timelike particle is exactly proportional to the period $T_{r}$ of the radial coordinate, such as $T_{\varphi}/T_{r}=q$, where $q$ is an arbitrary rational number.

The authors in \cite{2008PhRvD..77j3005L} point out that the orbital rational number $q$ can reflect the orbital characteristics from two aspects. On one hand, it is defined as
\begin{equation}\label{13}
q=w+\frac{v}{z}.
\end{equation}
Here, $z$ represents the number of leaves the orbit can exhibit within a complete orbital period, which can also be understood as the number of times the timelike particle passes through the apastron. $w$ counts the nearly circular loops the orbit completes around the black hole near the periastron. In the equatorial plane described by the Cartesian coordinate system $(x,y)$, a larger $w$ implies that there are more intersection points between the orbit and the $x$-axis within a complete orbital period. $v$ is used to indicate the direction of the next apastron that the timelike particle heads towards after being released from the current apastron. Usually, along the direction of increasing $\varphi$, the apastrons are sequentially labeled as $v=0,1,2,3...z-1$. Specifically, for a single leaf orbit ($z=1$), $v$ can take the value of $0$ or $1$. Overall, once $(z,w,v)$ is given, the orbital configuration is determined and corresponds to the rational number $q$.

On the other hand, during the process where the orbit evolves from the apastron to the periastron and then reaches the next adjacent apastron, the accumulation of the azimuthal angle $\Delta\varphi$ is related to the rational number $q$:
\begin{eqnarray}\label{14}
q = \frac{\Delta\varphi}{2\pi} - 1.
\end{eqnarray}
According to the geodesic integration, we have
\begin{eqnarray}\label{15}
\Delta\varphi = 2\int^{\varphi_{2}}_{\varphi_{1}}\textrm{d}\varphi = 2\int^{r_{2}}_{r_{1}}\frac{\dot{\varphi}}{\dot{r}}\textrm{d}r,
\end{eqnarray}
where $\varphi_{1}$ and $\varphi_{2}$ represent the azimuthal angles of the periastron and apastron respectively, and the derivatives are written as
\begin{eqnarray}
\dot{\varphi} = \frac{L_{i}}{g_{\varphi\varphi}}, \label{16} \\
\dot{r} = \sqrt{\frac{-E_{i}^{2}}{g_{tt}g_{rr}}-\frac{1}{g_{rr}}\left(1+\frac{L_{i}^{2}}{g_{\varphi\varphi}}\right)}. \label{17}
\end{eqnarray}
By combining equations \eqref{14} and \eqref{15}, we establish the connection among the rational orbital configuration $(z,w,v)$ and the constants of motion $L_{i}$ and $E_{i}$. In addition, we can also obtain the proper time consumption of the rational orbit between adjacent periastron and apastron
\begin{equation}\label{18}
\Delta\tau = \int^{r_{2}}_{r_{1}}\frac{1}{\dot{r}}\textrm{d}r.
\end{equation}
Then, the period corresponding to a complete orbit is written as $T_{q} = 2z\Delta\tau$.

With the theoretical framework described above, the procedure for simulating rational timelike orbits in spherically symmetric spacetimes is highly clear. First, based on the conditions for the ISCO and the MBO, we calculate $L_{\textrm{isco}}$ and $L_{\textrm{mbo}}$. Second, we select $L_{i}$ within the open interval $(L_{\textrm{isco}}, L_{\textrm{mbo}})$, use equation \eqref{11} to calculate the radii of the stable and unstable circular orbits corresponding to this angular momentum, and obtain the specific energies corresponding to these two radii according to equation \eqref{9}. Thus, we acquire the specific energy range $(E_{\textrm{min}},E_{\textrm{max}})$ for stable precessing orbits for a given $L_{i}$. Third, by scanning $E_{i}$ within $(E_{\textrm{min}},E_{\textrm{max}})$, we calculate the rational number $q$ corresponding to $(L_{i},E_{i})$ according to equation \eqref{15}, and obtain the mapping relationship between $q$ and $E_{i}$. Fourth, given the configuration parameters $(z,w,v)$ of the target orbit, we utilize equation \eqref{13} to calculate the corresponding rational number $q$, and employ interpolation methods to obtain the $E_{i}$. By substituting $E_{i}$ and $L_{i}$ into equation \eqref{9}, we obtain the periastron and apastron radii $r_{1}$ and $r_{2}$. Consequently, the initial conditions $(t,r,\theta,\varphi)$ and $(p_{t},p_{r},p_{\theta},p_{\varphi})$ for the rational orbit corresponding to $(z,w,v)$ are read as $(0,r_{2},\pi/2,0)$ and $(-E_{i},0,0,L_{i})$. Finally, by utilizing a numerical integration algorithm to integrate the canonical equations \eqref{8} over the proper time $T_{q}$, we can obtain the evolution of the target rational orbit within a complete period.

It is not difficult to deduce that the key to obtaining a rational orbit lies in confirming the precise values of the orbital specific energy and specific angular momentum corresponding to $(z,w,v)$. If these values deviate, the closed nature of the orbit is broken, and the rational orbit transitions into an irrational orbit. To obtain the irrational orbit, one can add a minute perturbation to $q$. For instance, the $q=1+1/2$ corresponding to $(2,1,1)$ can be modified to $q=1+1/2+1/200=1+101/200$. Then, the modified $q$ is subjected to the aforementioned interpolation procedure to obtain the initial conditions of the irrational orbit. The irrational orbit simulated in this manner can still exhibit the $(z,w,v)$ characteristics over a short evolution time. However, as the evolution time increases, the envelope of the orbit eventually evolves into a circle.
\subsection{Gravitational Waves and Characteristic Strains}
Under general circumstances, small celestial bodies orbit around supermassive black holes at the centers of galaxies, and the mass ratio between them can exceed $10^{4}$. The orbits of these small celestial bodies can be characterized by geodesics and inherently encode spacetime information. More importantly, the extreme-mass-ratio inspiral systems composed of small celestial bodies and central black holes serve as the observational targets for space gravitational wave projects. To this end, our code further integrates the computation of gravitational wave waveforms and characteristic strains radiated by precessing orbits, including both rational and irrational orbits.

In \cite{2007PhRvD..75b4005B}, the authors provided approximate analytical formulas for calculating the gravitational wave polarization states of extreme-mass-ratio inspiral systems, which are sufficient to capture the essential gravitational wave features, written as
\begin{eqnarray}
h_{+} = -\frac{2\eta}{D_{\textrm{L}}}\frac{GM}{c^{2}r}\left(1+\cos^{2}\iota\right)\cos\left(2\varphi+2\zeta\right), \label{19} \\
h_{\times} = -\frac{4\eta}{D_{\textrm{L}}}\frac{GM}{c^{2}r}\cos\iota\sin\left(2\varphi+2\zeta\right). \label{20}
\end{eqnarray}
Here, $\eta$ is the symmetric mass ratio parameter of the system, given by $\eta=mM/(m+M)^{2}$, where $m$ is the mass of the small celestial body. $D_{\textrm{L}}$ represents the luminosity distance from the extreme-mass-ratio inspiral system to the Earth. $\iota$ and $\zeta$ denote the inclination angle and the longitude of the periastron of the orbit, respectively. After obtaining $r$ and $\varphi$ at any given moment of the orbit, substituting them into equations \eqref{19} and \eqref{20} allows the time-domain information of the two gravitational wave polarization states to be obtained directly.

Based on the fast Fourier transform, we can convert the gravitational wave signals calculated from equations \eqref{19} and \eqref{20} into the frequency-domain, yielding $\hat{h}_{+}\left(f\right)$ and $\hat{h}_{\times}\left(f\right)$, where $f$ represents the frequency. Furthermore, by combining the frequency-domain signals of the two polarization states according to \cite{Finn:2000sy}
\begin{equation}\label{21}
h_{\textrm{c}}\left(f\right) = 2f\sqrt{|\hat{h}_{+}\left(f\right)|^{2}+|\hat{h}_{\times}\left(f\right)|^{2}},
\end{equation}
we can obtain the characteristic strain $h_{\textrm{c}}$ of the gravitational wave. By comparing the $h_{\textrm{c}}$ of different orbits with the sensitivity curves of space gravitational wave detectors, the observational prospects of the orbital gravitational radiation can be thoroughly discussed.
\section{Demonstration of Code Capabilities in the Schwarzschild Spacetime}
In geometrized units, the covariant metric of the Schwarzschild spacetime is written as
\begin{eqnarray}\label{22}
g_{tt}=-\left(1-\frac{2}{r}\right), \quad g_{rr}=\frac{-1}{g_{tt}}, \quad g_{\theta\theta}=r^{2}, \quad g_{\varphi\varphi}=r^{2}\sin^{2}\theta.
\end{eqnarray}
Our code requires only the definition of the covariant metric to operate with full functionality.
\begin{figure*}
\centering                   
\includegraphics[width=6cm]{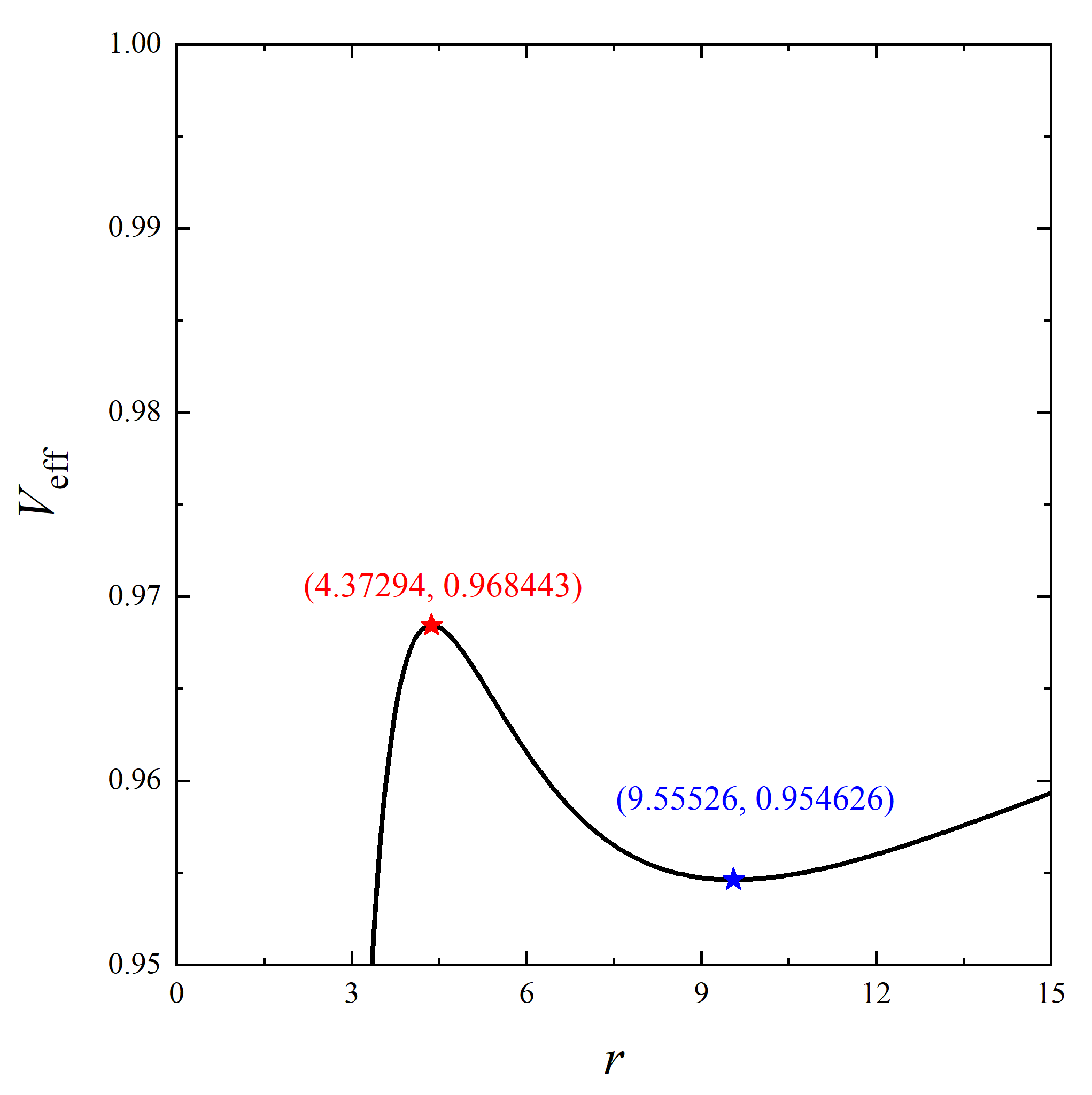}
\caption{Effective potential $V_{\textrm{eff}}$ as a function of $r$ in the case of $L_{i}=3.7321$. The red and blue star markers correspond to the maximum and minimum points respectively, representing the locations of the unstable circular orbit and the stable circular orbit.}      
\label{fig1}                 
\end{figure*}

According to the conditions for the ISCO and the MBO, we obtain $r_{\textrm{isco}}=6$ and $r_{\textrm{mbo}}=4$ for the Schwarzschild spacetime, with the corresponding orbital angular momenta being $L_{\textrm{isco}}=3.4641$ and $L_{\textrm{mbo}}=4$. To quantify the selection of $L_{i}$, we introduce the angular momentum parameter $\varepsilon \in (0,1)$, such that 
\begin{equation}\label{23}
L_{i}=L_{\textrm{isco}}+\varepsilon\left(L_{\textrm{mbo}}-L_{\textrm{isco}}\right).
\end{equation}
We take $\varepsilon=0.5$, which corresponds to $L_{i}=3.7321$, as an example to demonstrate the results. Figure \ref{fig1} displays the evolution of the effective potential $V_{\textrm{eff}}$ with respect to the radial coordinate $r$. It can be observed that the effective potential curve exhibits two extremum points, where the maximum point corresponds to the unstable circular orbit and the minimum point corresponds to the stable circular orbit. Their radii can be obtained by solving equation \eqref{11}, yielding $4.37294$ and $9.55526$ respectively. According to equation \eqref{9}, we further determine the specific energies corresponding to the maximum and minimum values to be $E_{\textrm{max}}=0.968443$ and $E_{\textrm{min}}=0.954626$, respectively. In other words, for $L_{i}=3.7321$, there exists a specific energy range $(E_{\textrm{min}},E_{\textrm{max}})$ that allows stable precessing orbits to exist.
\begin{figure*}
\centering                   
\includegraphics[width=6cm]{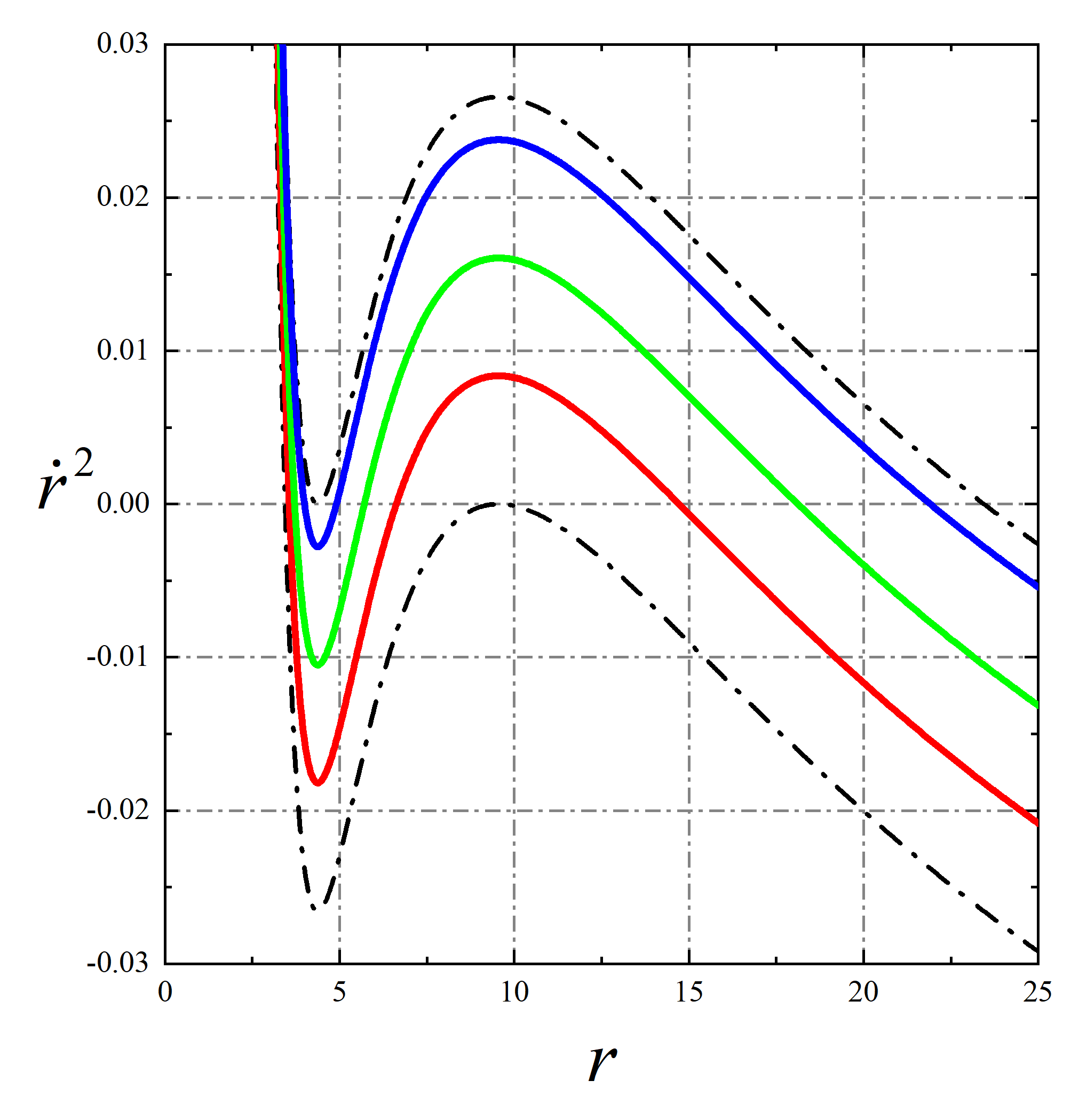}
\caption{$\dot{r}^{2}$ as a function of $r$. The red, green, and blue solid lines correspond to $E_{1}=0.959$, $E_{2}=0.963$, and $E_{3}=0.967$ respectively. The upper and lower dash dotted lines correspond to $E_{\textrm{max}}$ and $E_{\textrm{min}}$ respectively. It can be observed that the red, blue, and green lines intersect with $\dot{r}^{2}=0$ three times, indicating the existence of stable orbits, and the two larger roots correspond to the apastron and periastron of the orbits. In contrast, the two dash-dotted lines intersect with $\dot{r}^{2}=0$ only twice, corresponding to circular orbits.}   
\label{fig2}                 
\end{figure*}

We select $E_{1}=0.959$, $E_{2}=0.963$, and $E_{3}=0.967$ within the specific energy range and plot the variation of $\dot{r}^{2}$ with respect to $r$ respectively, as illustrated in figure \ref{fig2}, where the black dash-dotted lines are contributed by $E_{\textrm{min}}$ and $E_{\textrm{max}}$. We find that the curves corresponding to the three different specific energies $E_{i}$ are completely located between the two critical dash-dotted lines, and they intersect with $\dot{r}^{2}=0$ three times, indicating the existence of precessing orbits. The timelike particles can perform oscillatory motion within the lobes of their respective curves where $\dot{r}^{2}>0$ (between the two larger roots of $\dot{r}^{2}=0$). This result not only demonstrates the high precision of the code but also verifies the feasibility of the previously mentioned method for finding stable precessing orbits.
\begin{figure*}
\centering                   
\includegraphics[width=6cm]{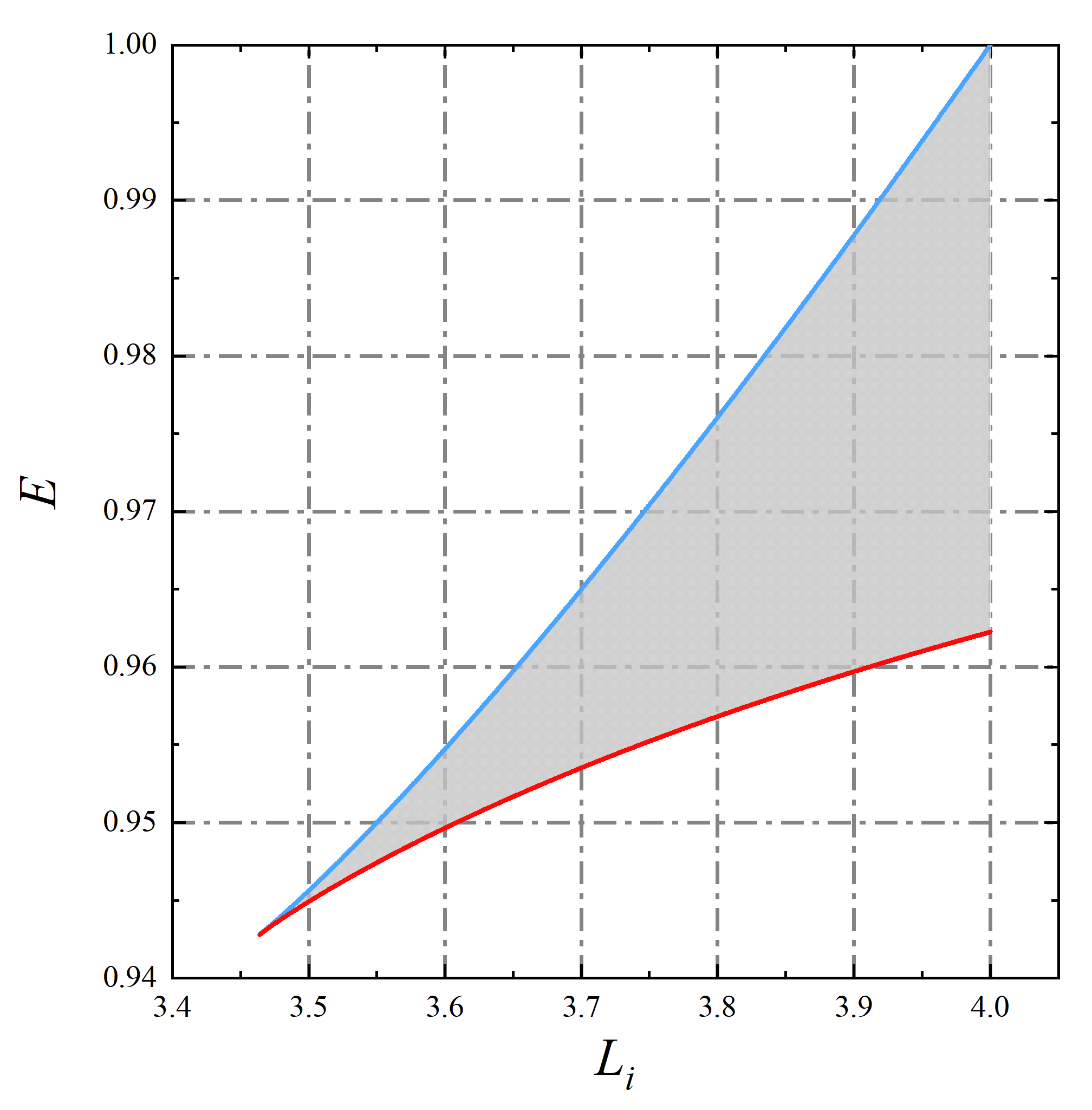}
\caption{Evolution of the maximum specific energy $E_{\textrm{max}}$ and the minimum specific energy $E_{\textrm{min}}$ corresponding to stable orbits with respect to the orbital angular momentum $L_{i}$. The shaded region between the two curves represents the parameter space where stable orbits exist.}     
\label{fig3}                 
\end{figure*}

Our code also supports searching for the conditions of stable precessing orbits in a broader parameter space. Figure \ref{fig3} illustrates the evolution of $E_{\textrm{min}}$ (red solid line) and $E_{\textrm{max}}$ (blue solid line) with respect to the specific angular momentum $L_{i}$. It can be found that both critical values of $E$ increase as $L_{i}$ increases, and the region between them resembles a triangle. Arbitrarily selecting a combination of specific energy and specific angular momentum within the shaded region yields a stable bound orbit.
\begin{figure*}
\centering                   
\includegraphics[width=7cm]{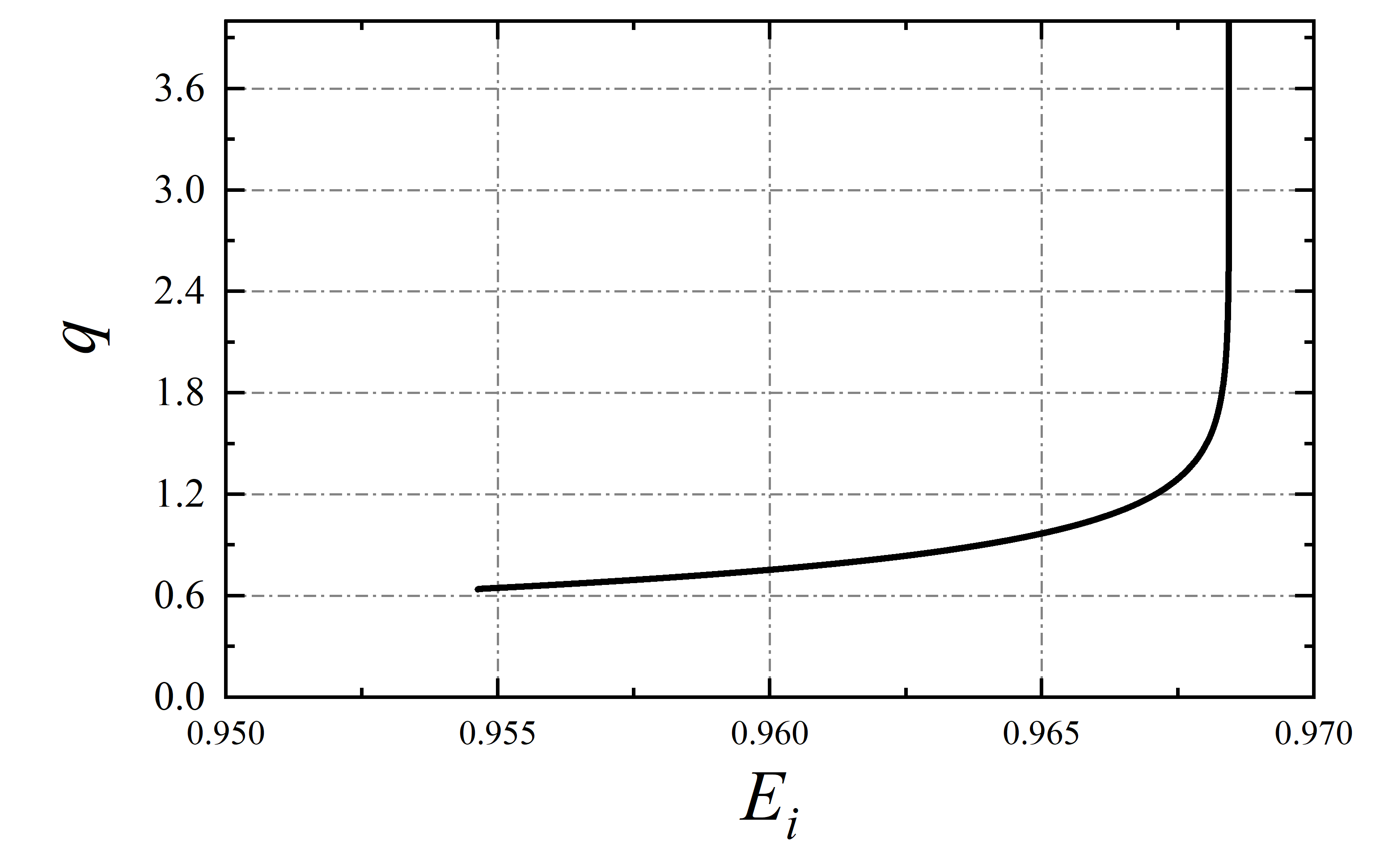}
\caption{Rational number $q$ as a function of the specific energy $E_{i}$. It is clear that $q$ exhibits a lower limit since $E_{i}$ possesses a minimum value.}      
\label{fig4}                 
\end{figure*}

Next, we return to the case of $\varepsilon=0.5$. Figure \ref{fig4} illustrates the evolution of the rational number $q$ with respect to $E_{i}$, obtained by combining equations \eqref{14} and \eqref{15}, where $E_{i} \in (E_{\textrm{min}},E_{\textrm{max}})$. It can be observed that the variation of $q$ initially exhibits a slow increase as $E_{i}$ grows, and subsequently experiences a sharp increase when the specific energy reaches $E_{\textrm{max}}$. This occurs because when $E_{i}$ approaches its maximum value, the orbit tends to circularize, which naturally leads to a rapid accumulation in $\Delta\varphi$. Furthermore, we can discover that in this scenario, $q$ possesses a lower bound of approximately $q=0.637$ (the left endpoint of the curve). Combined with the connection between $q$ and the orbital configuration shown in equation \eqref{13}, this phenomenon indicates that the dynamics naturally filter out certain structures; for instance, the configuration $q=w+v/z=0+1/2$ is strictly prohibited.
\begin{figure*}
\centering                   
\includegraphics[width=3cm]{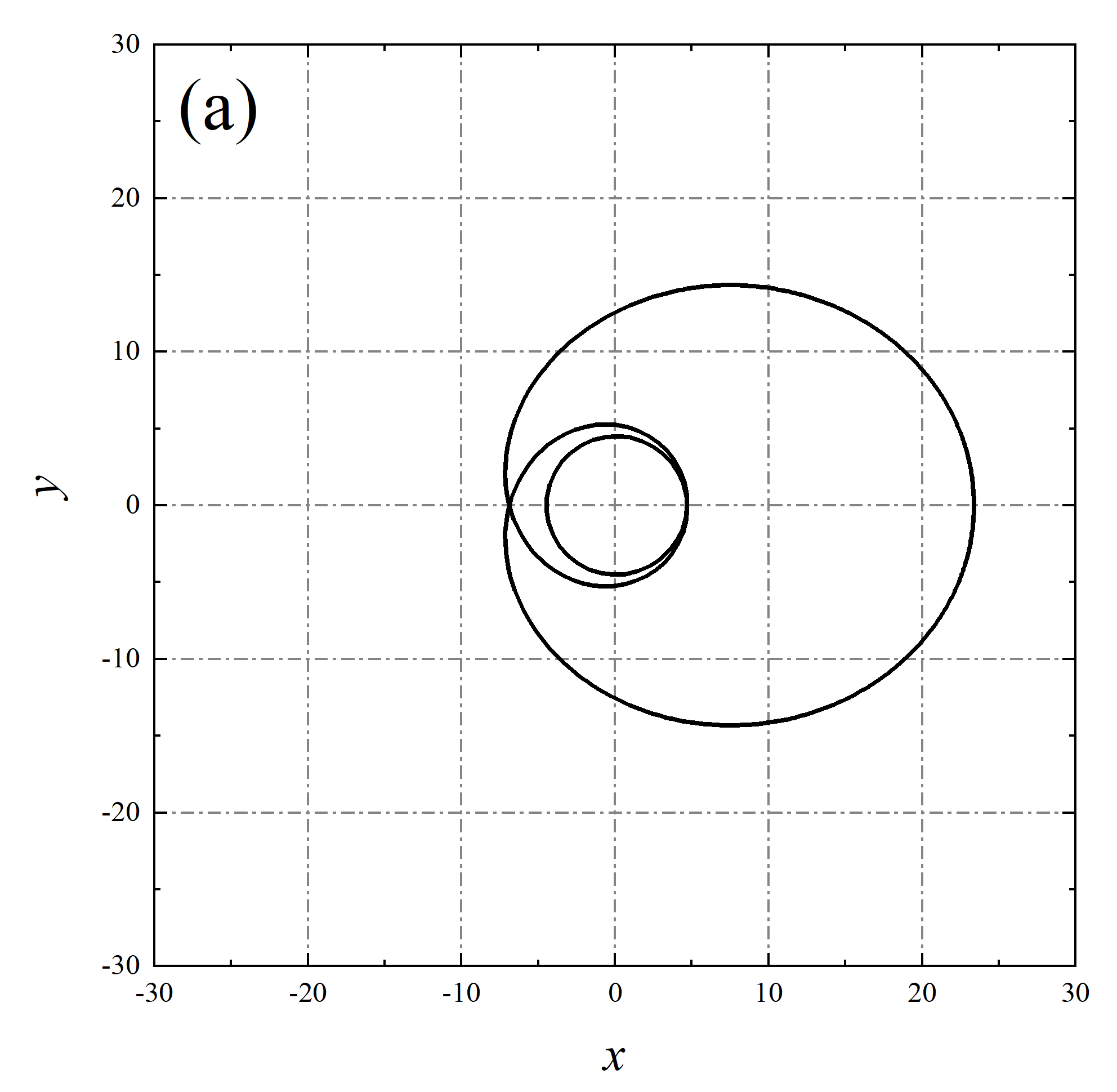}
\includegraphics[width=3cm]{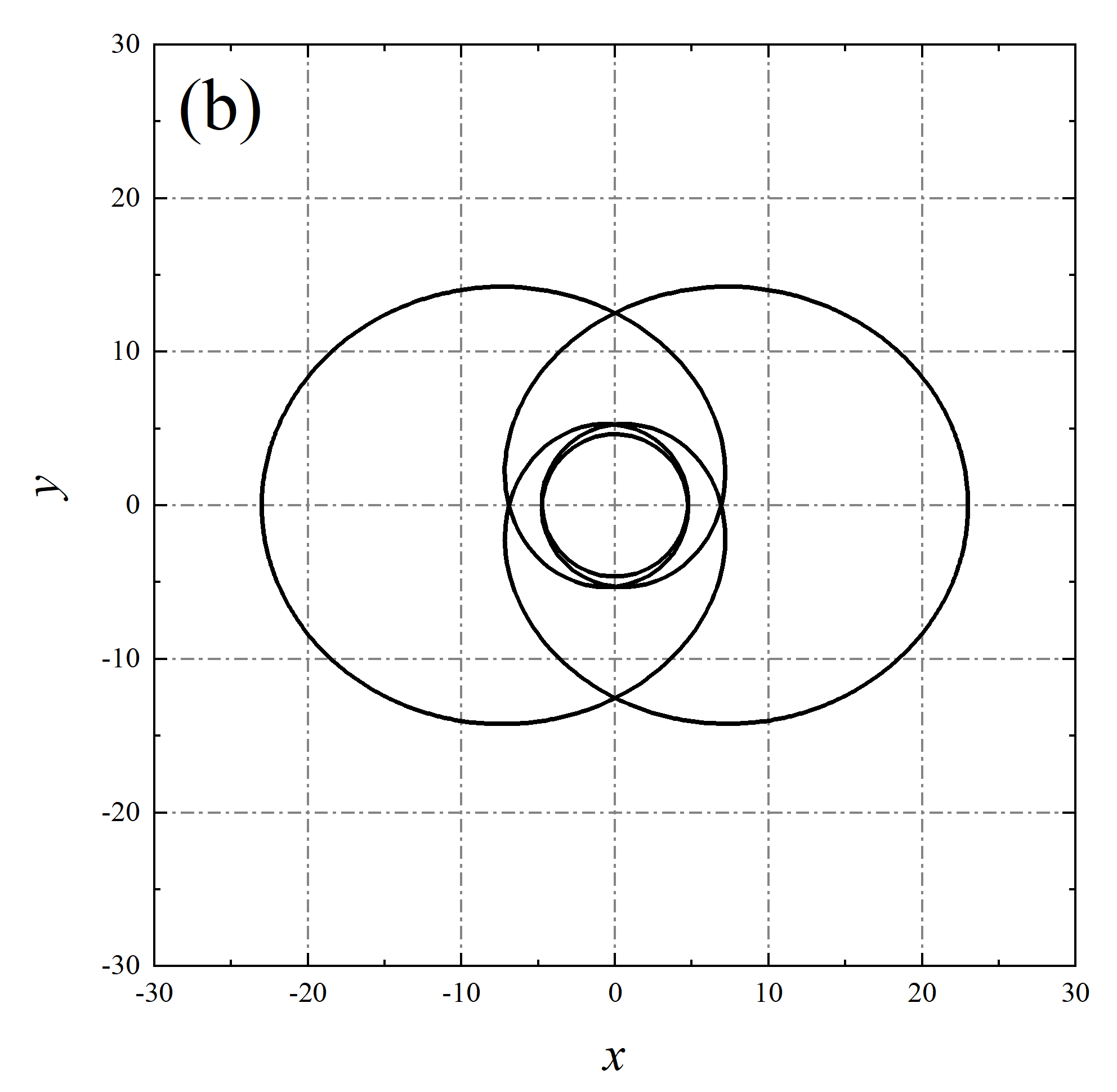}
\includegraphics[width=3cm]{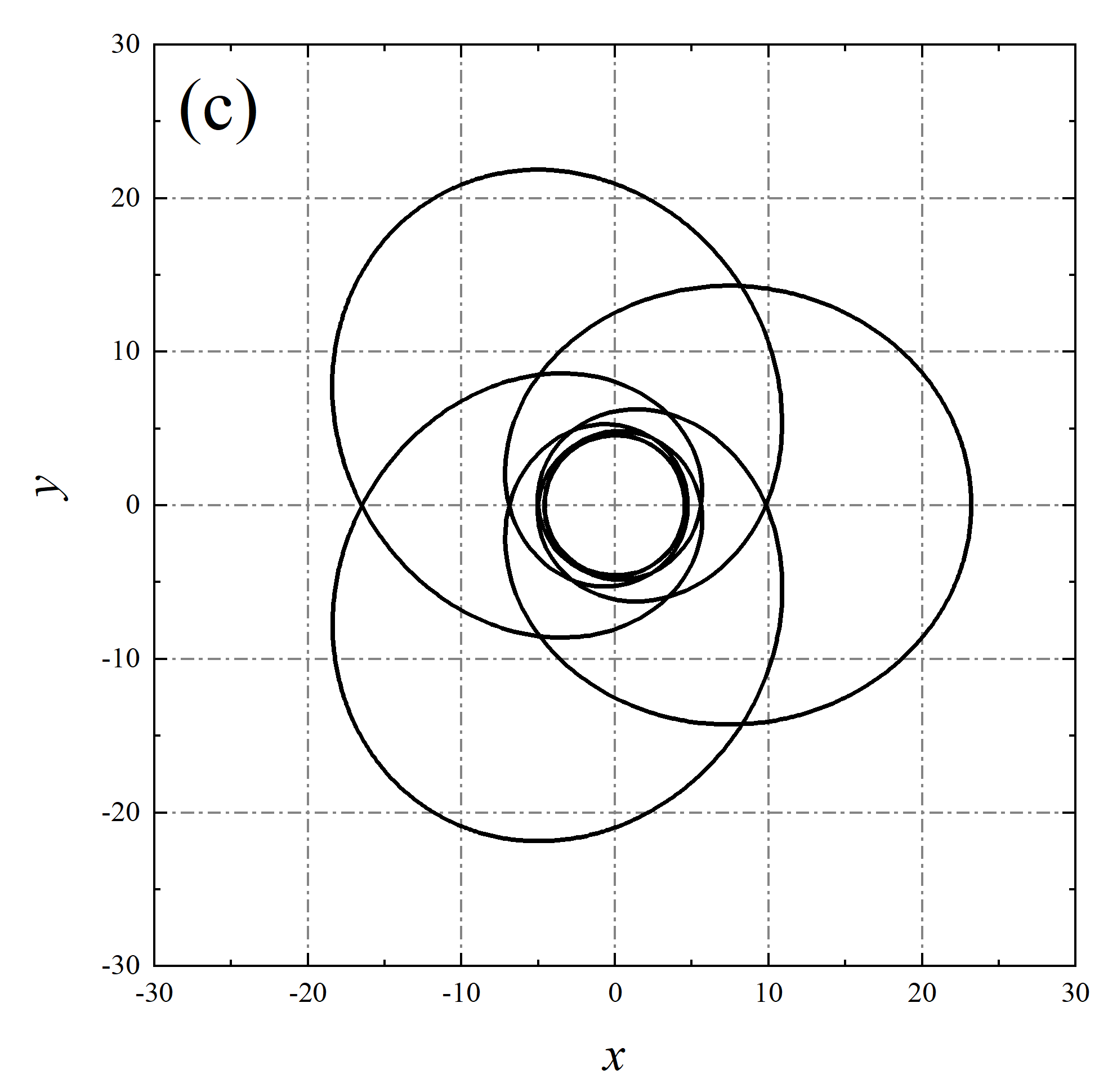}
\includegraphics[width=3cm]{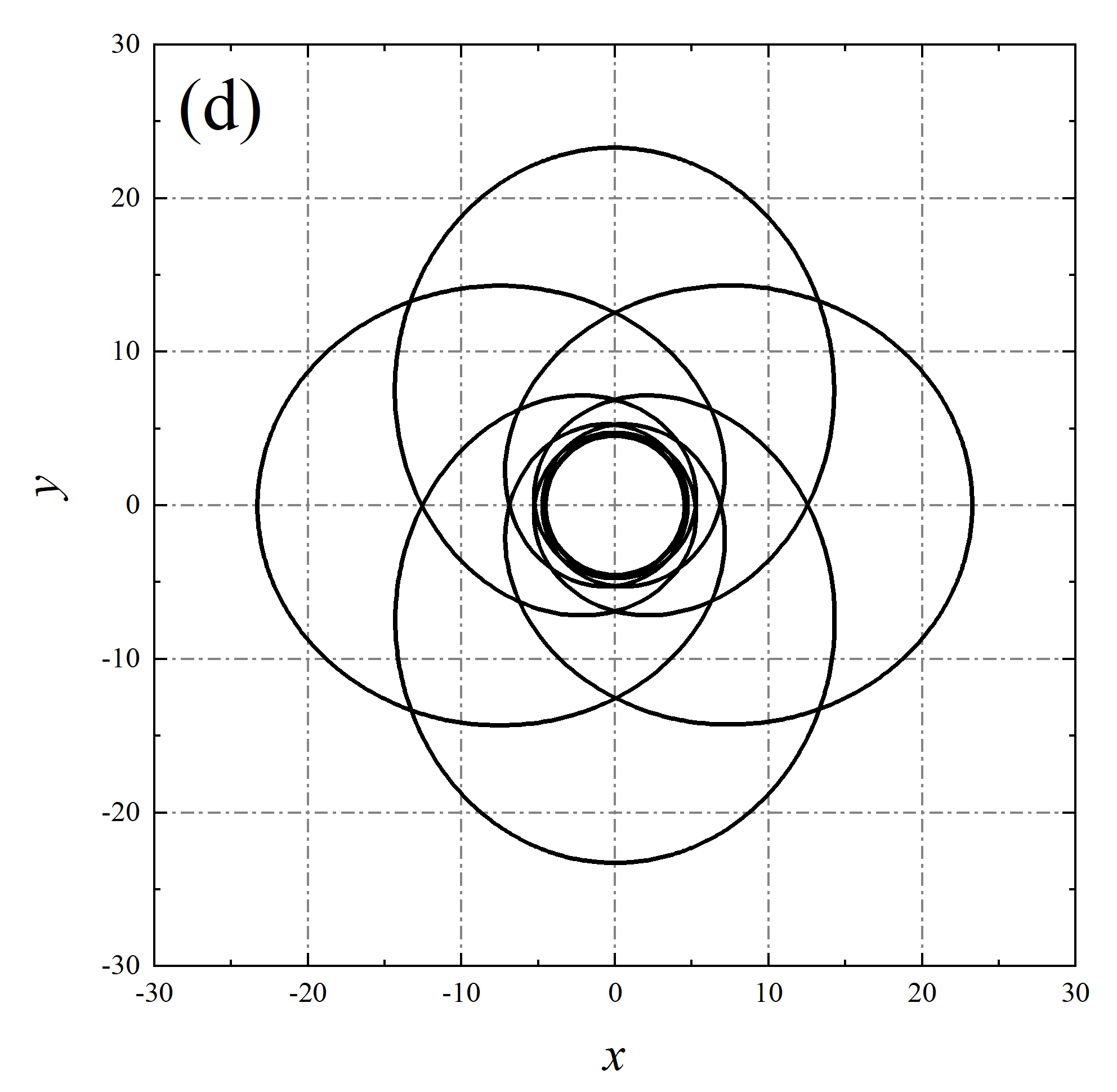}
\includegraphics[width=3cm]{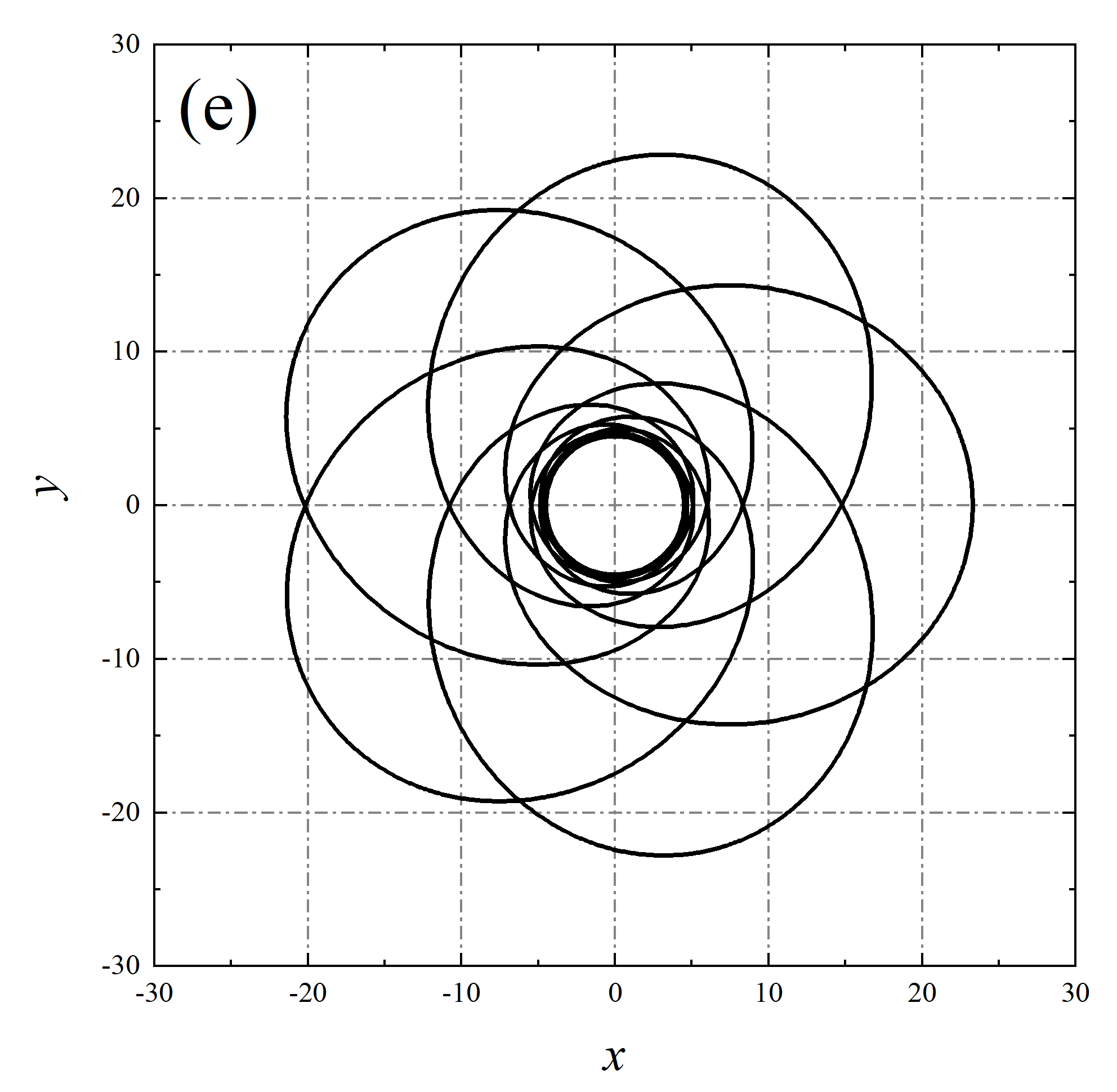}
\includegraphics[width=3cm]{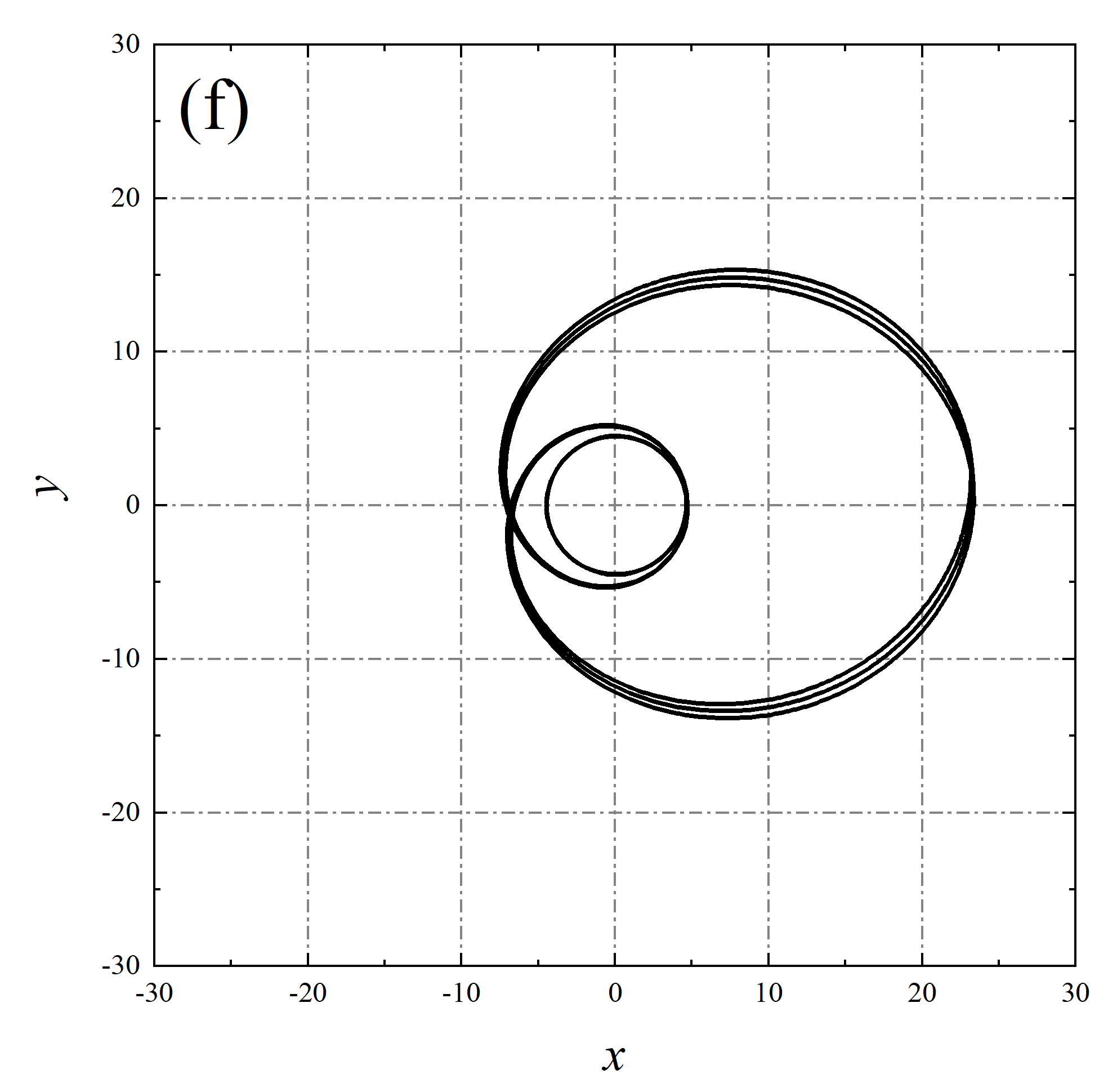}
\includegraphics[width=3cm]{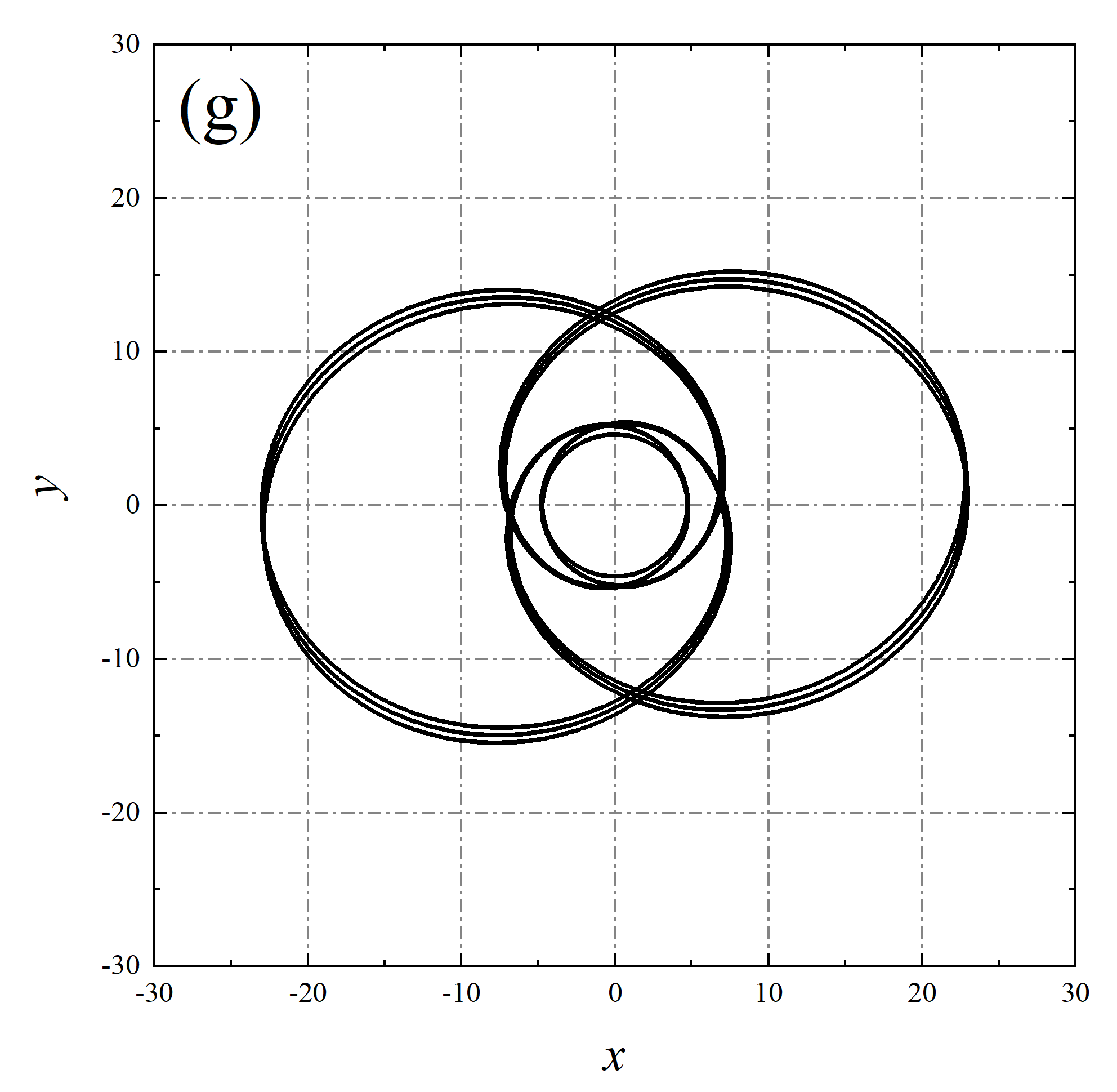}
\includegraphics[width=3cm]{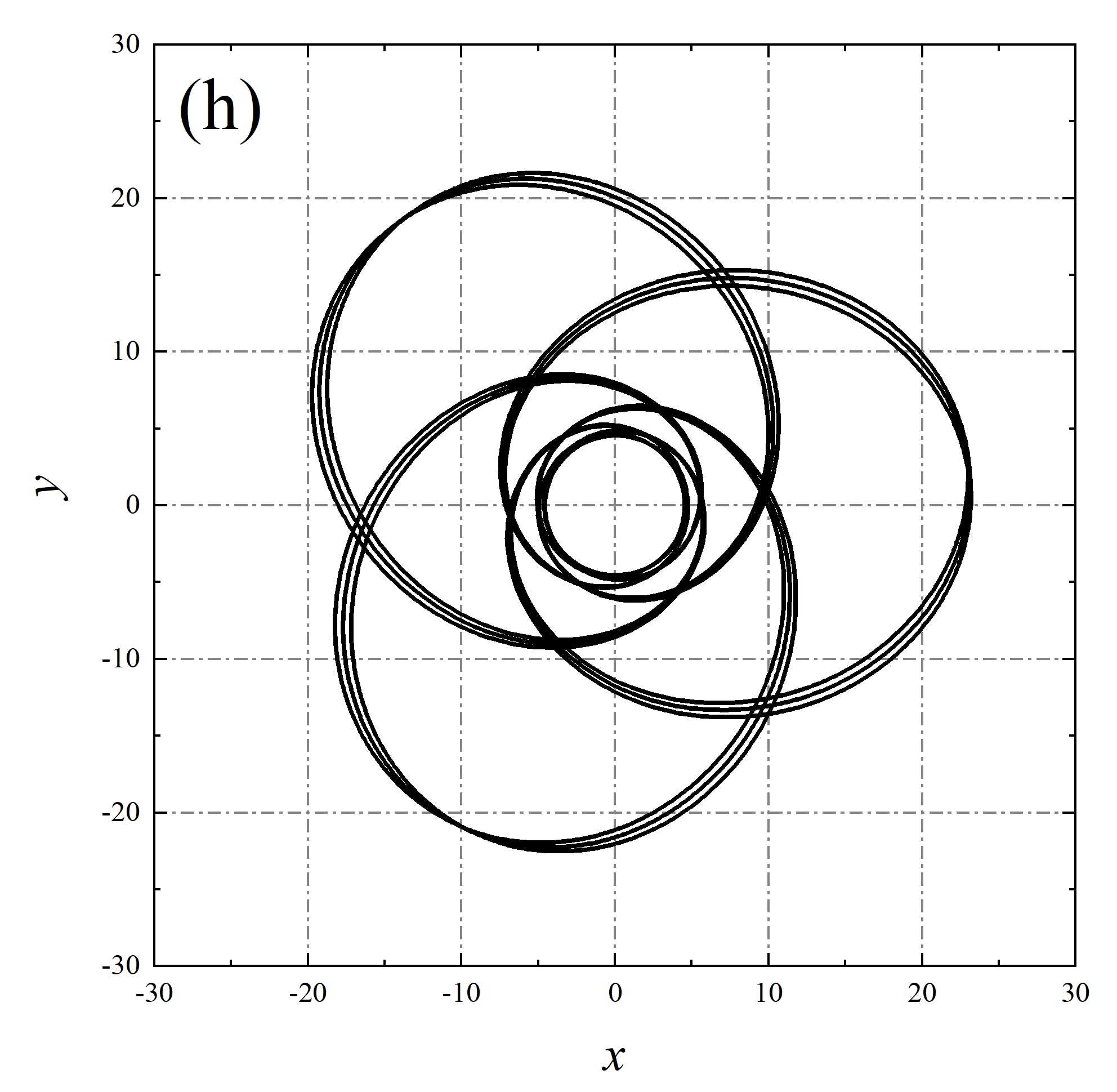}
\includegraphics[width=3cm]{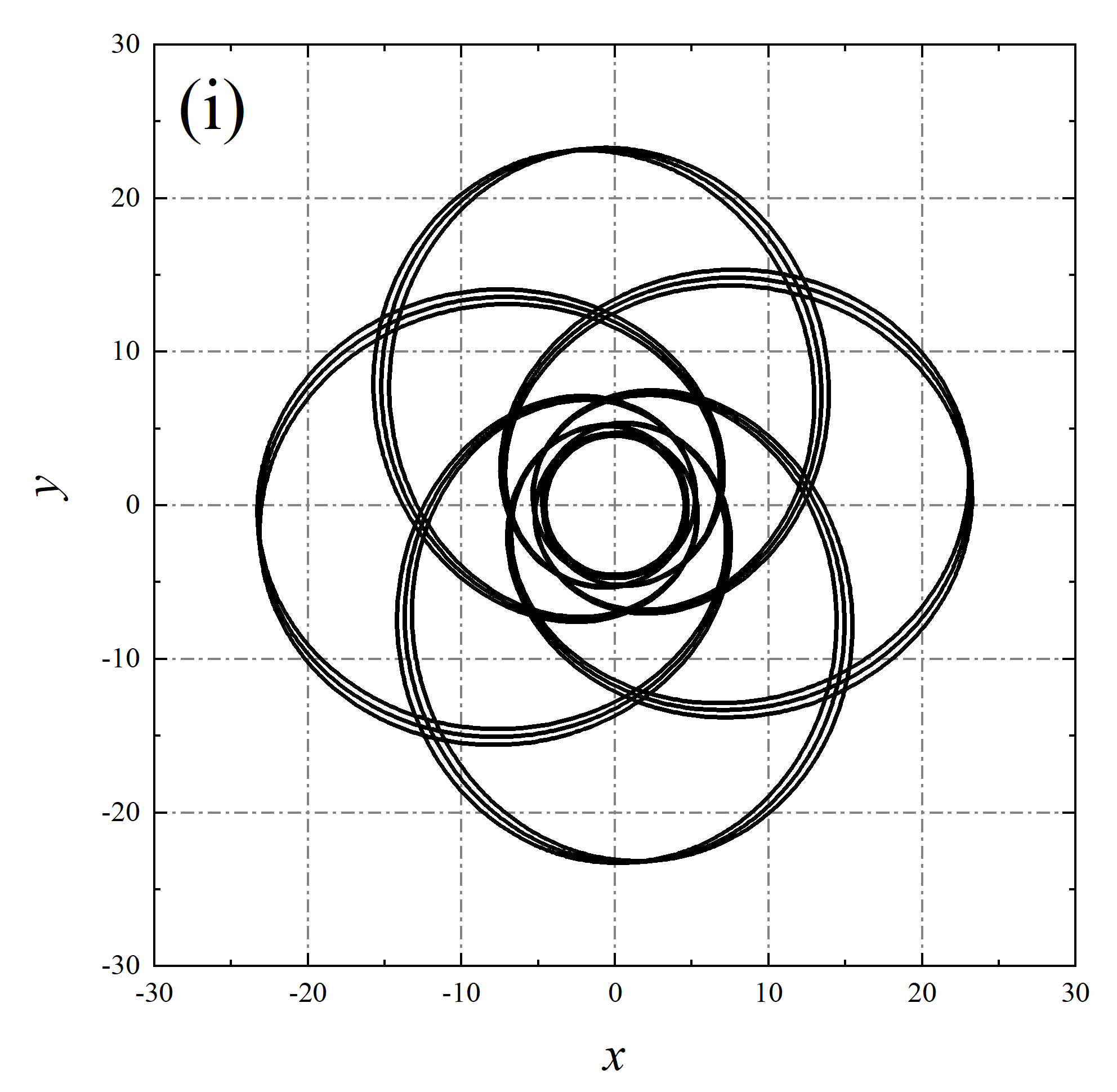}
\includegraphics[width=3cm]{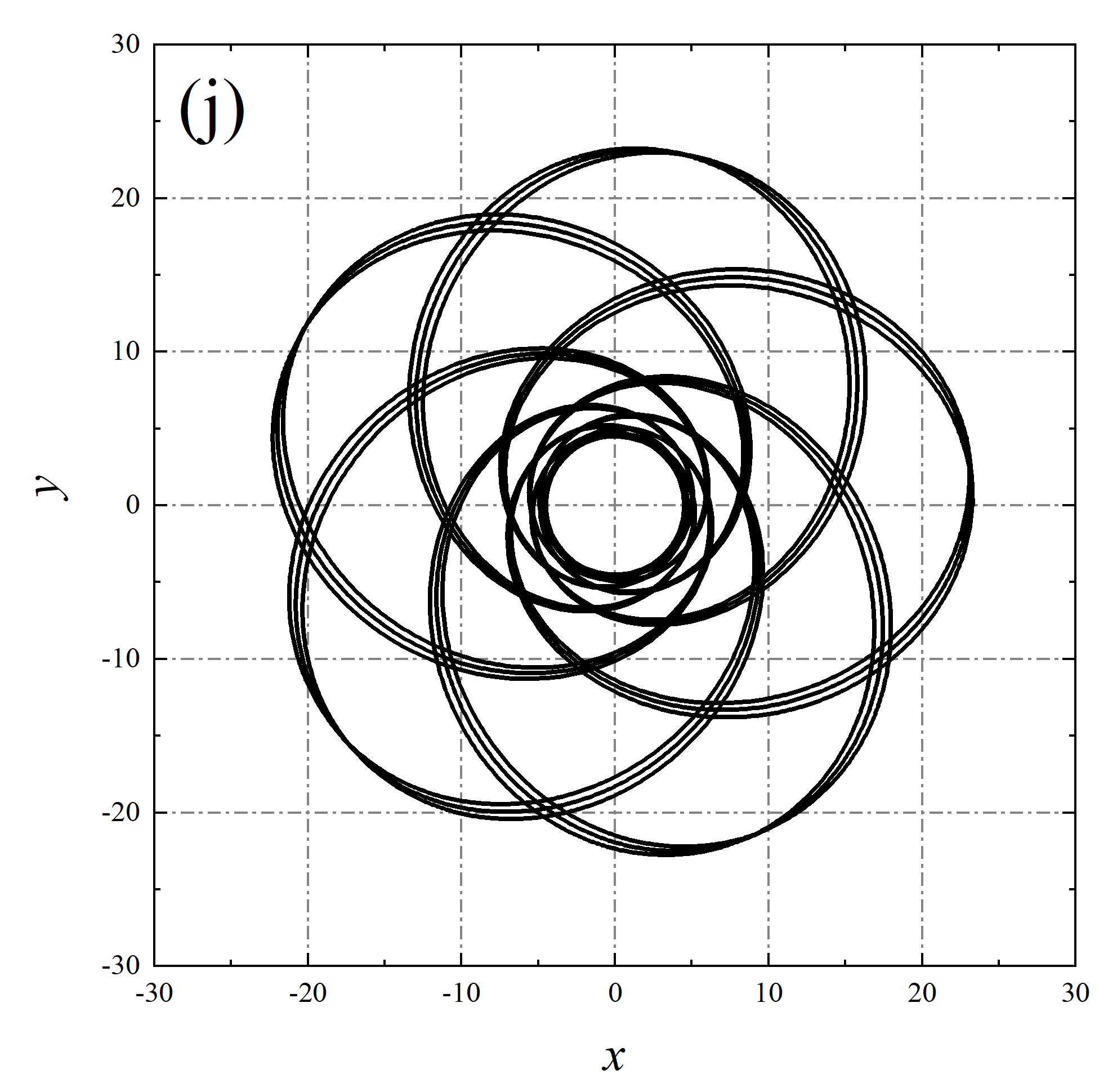}
\caption{Rational orbits (first row) and irrational orbits (second row) corresponding to different orbital configurations. From left to right, the number of orbital leaves are $z=1$, $2$, $3$, $4$, and $5$, and we have $v=z-1$ for $z > 1$, while $v=1$ for $z=1$. The irrational orbits are obtained by adding $1/100z$ to $q$. Here, we fix $w=1$. The simulated proper time for rational orbits is exactly one orbital period $T_{q}$, whereas the simulation time for irrational orbits is three times $T_{q}$.}       
\label{fig5}                 
\end{figure*}

Fixing $w=1$, we simulate the rational orbits for $z=1, 2, 3, 4, 5$ within their respective single periods, where $v=1$ for $z=1$, and $v=z-1$ for $z > 1$. As illustrated in the first row of figure \ref{fig5}, the starting point of each orbit is located at the rightmost apastron. It can be found that the number of orbital leaves, namely the number of apastrons, is perfectly consistent with $z$. The process of the orbit evolving from one apastron to the adjacent apastron contributes $w+1$ loops around the central black hole; therefore, as $z$ increases, the structure in the central region of the orbit becomes increasingly complex. Subsequently, we add a minute quantity to the rational number $q$ to observe the behavior of irrational orbits. As shown in the second row of figure \ref{fig5}, each orbit runs for three times the period of the original rational orbit. It can be observed that the irrational orbits lack strict periodicity and cannot perfectly overlap.
\begin{figure*}
\centering                   
\includegraphics[width=7.5cm]{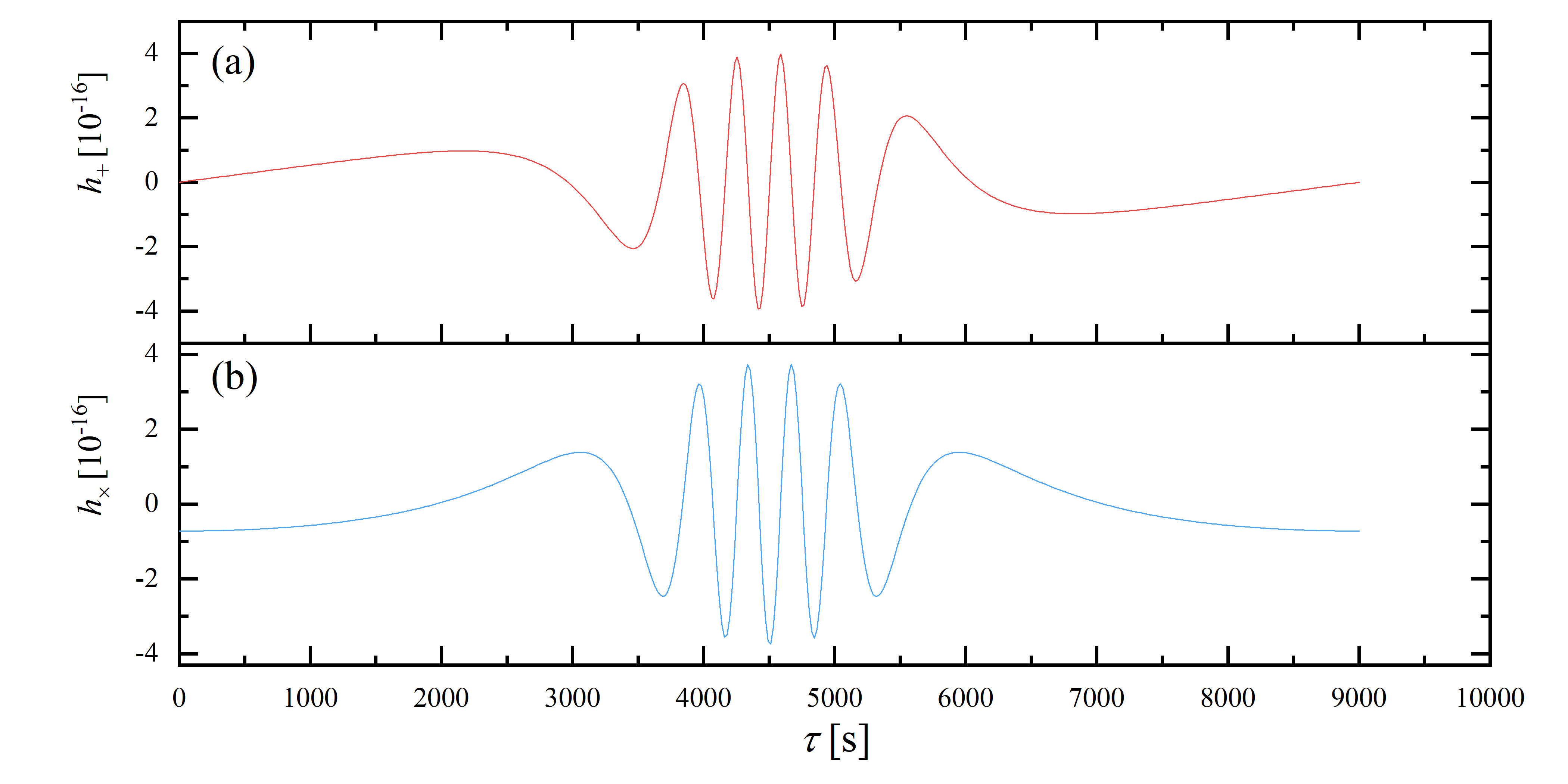}
\includegraphics[width=7.5cm]{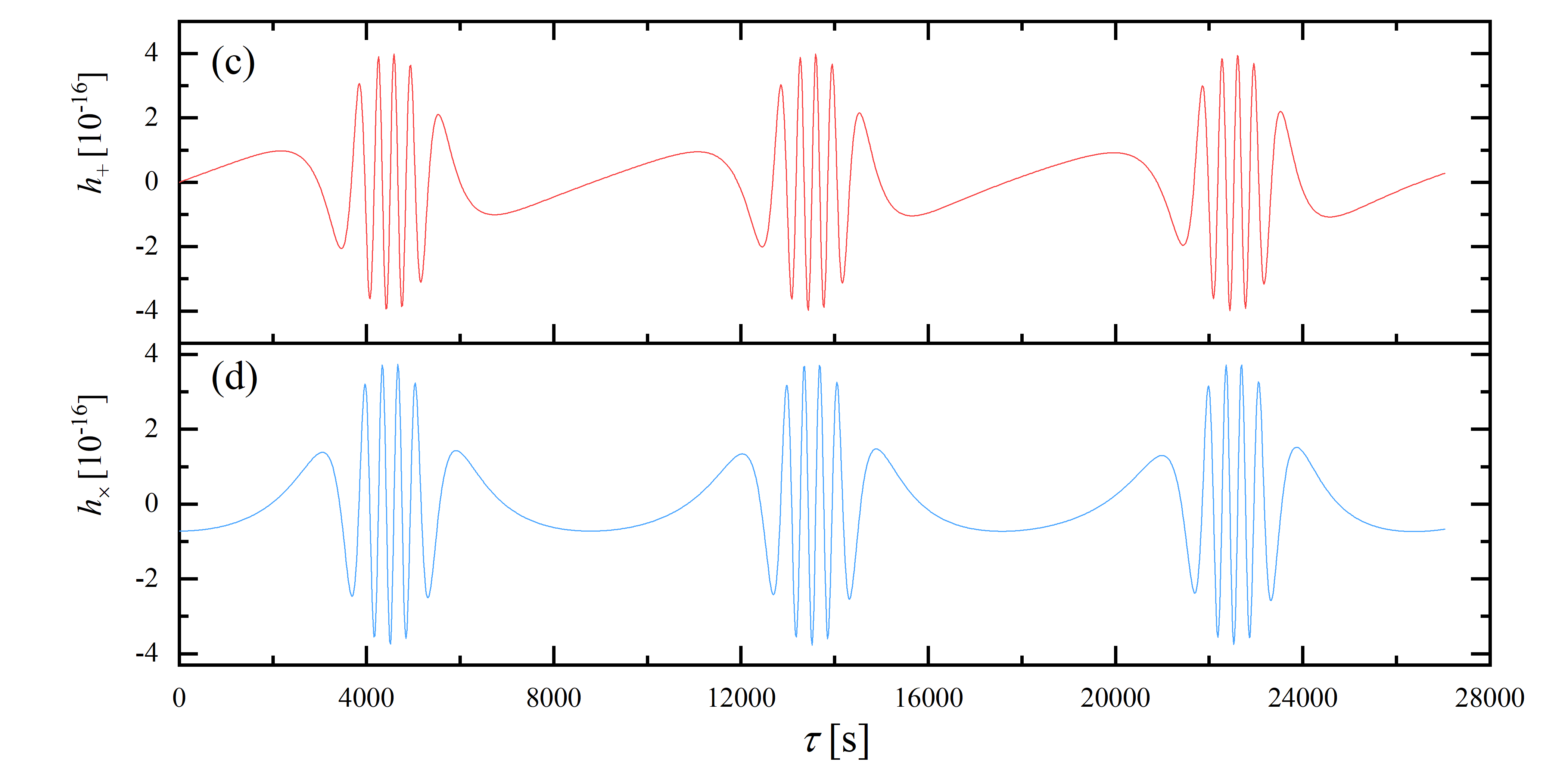}
\includegraphics[width=7.5cm]{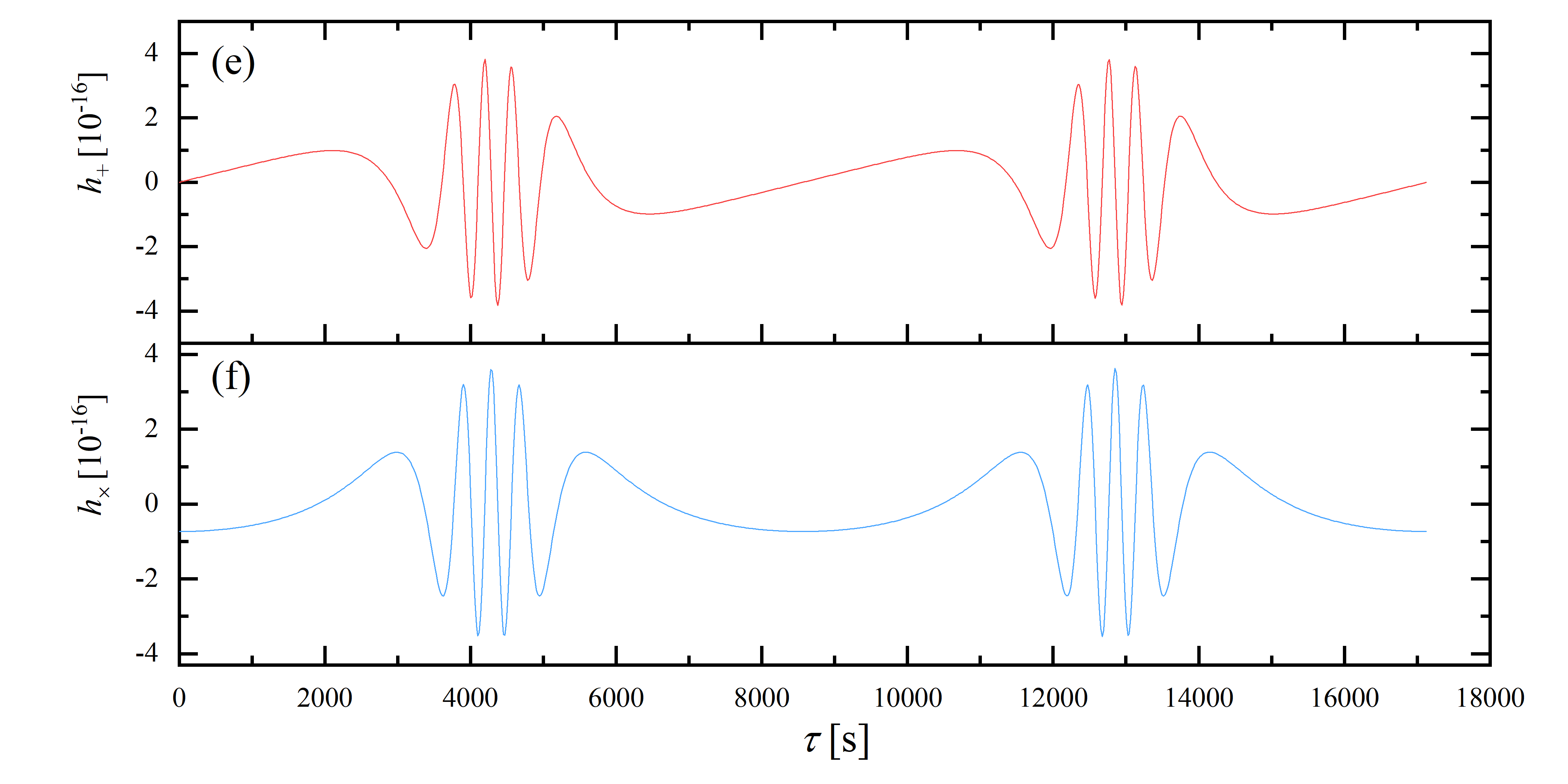}
\includegraphics[width=7.5cm]{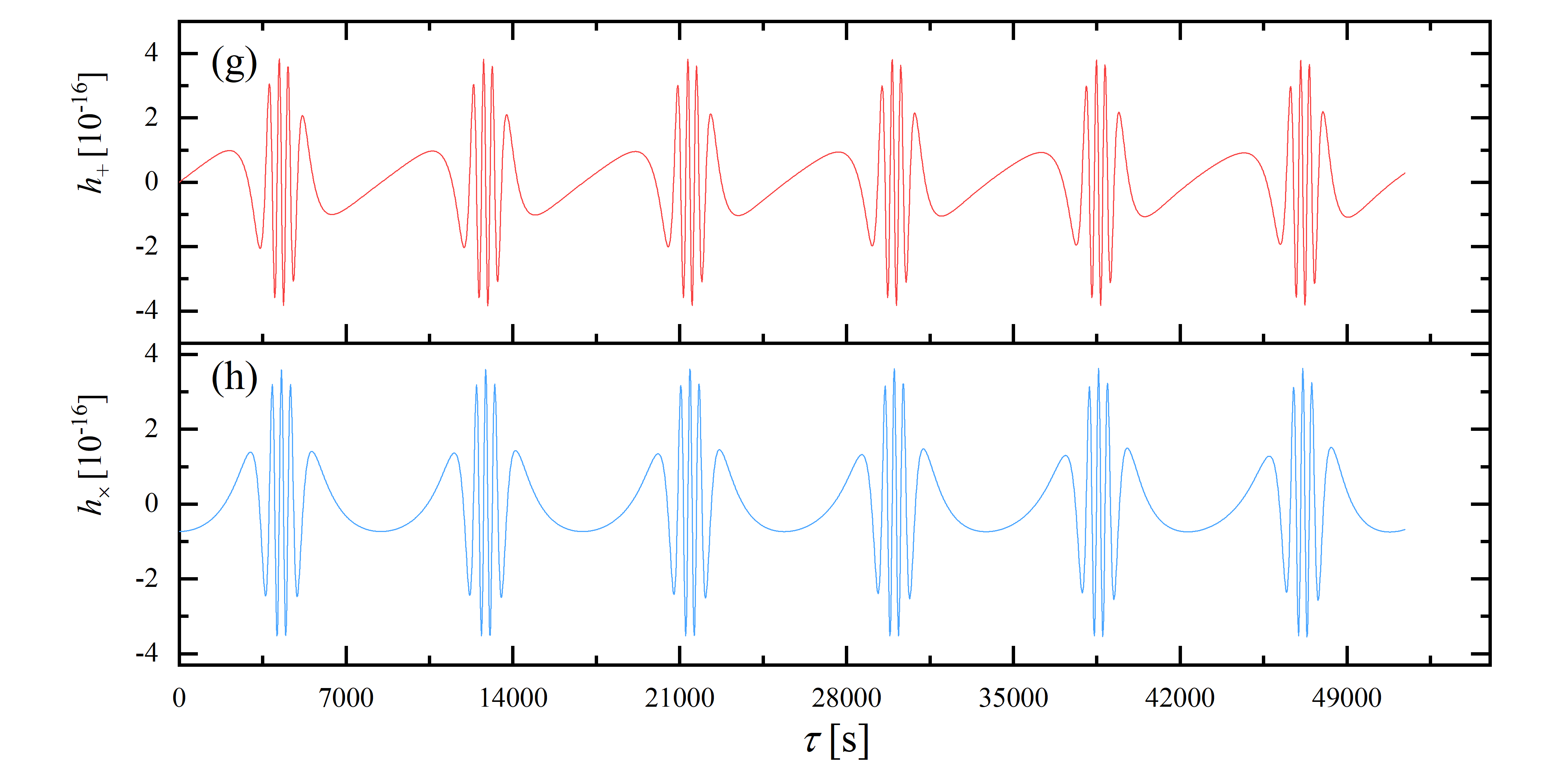}
\includegraphics[width=7.5cm]{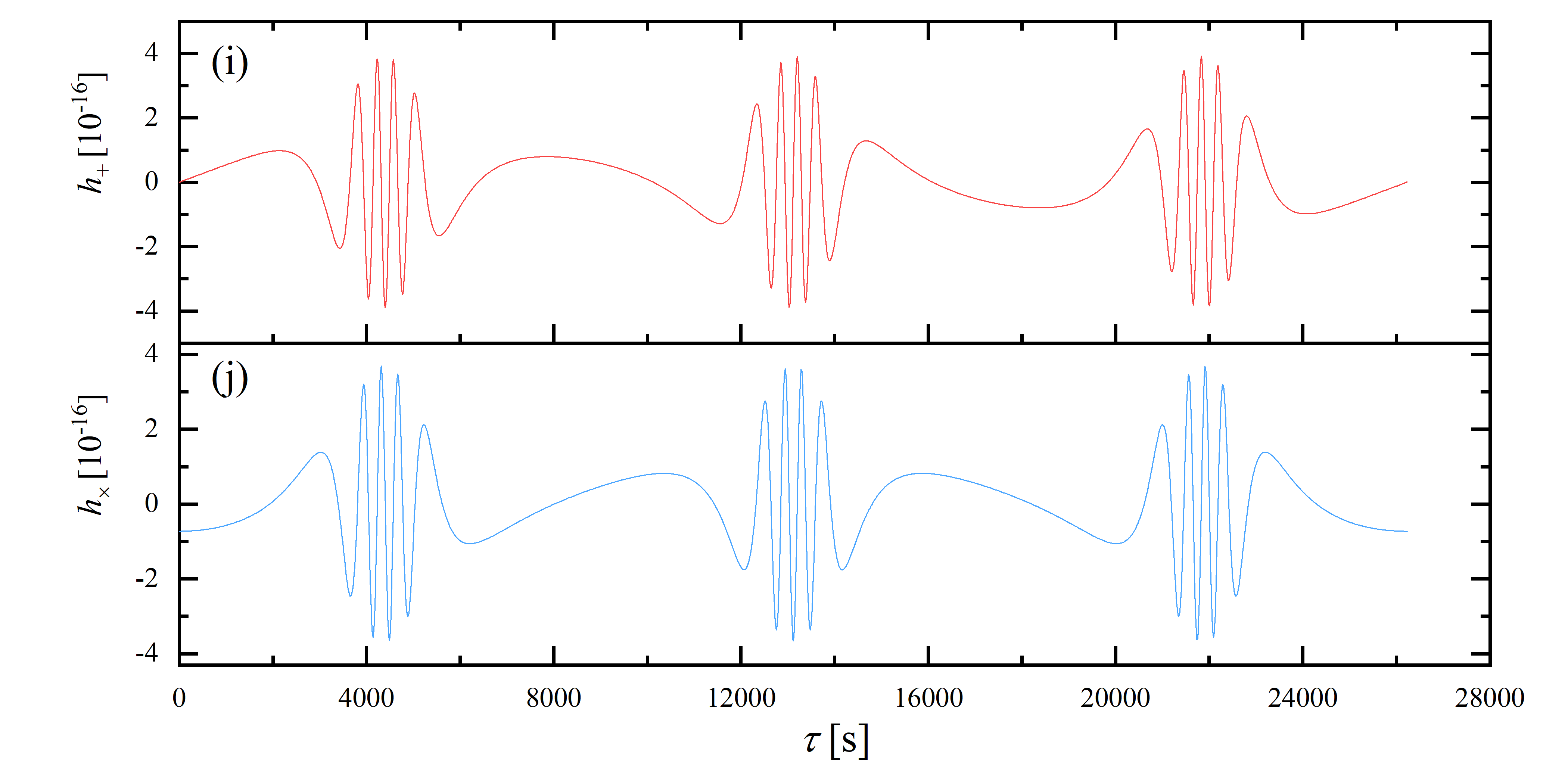}
\includegraphics[width=7.5cm]{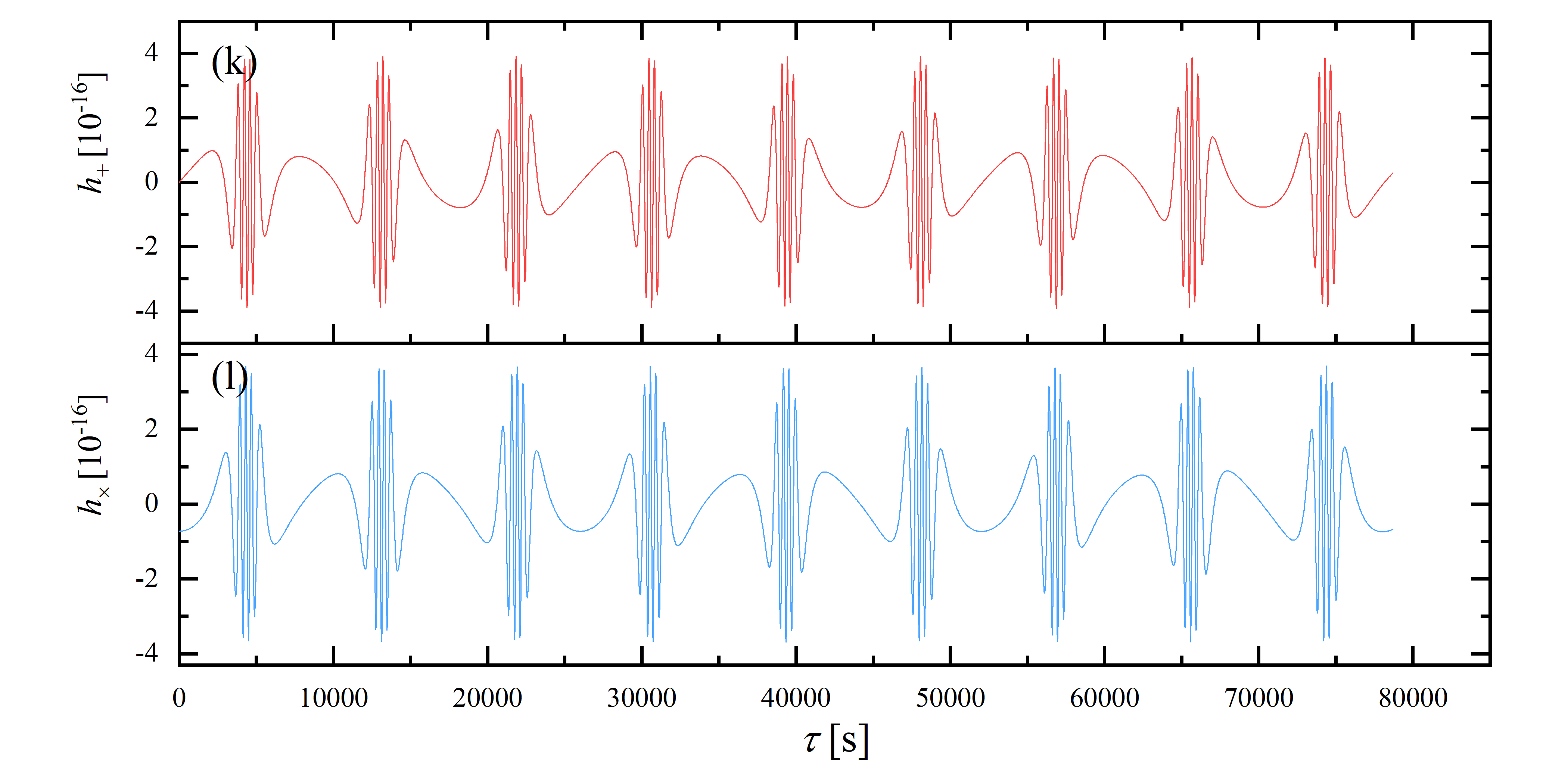}
\includegraphics[width=7.5cm]{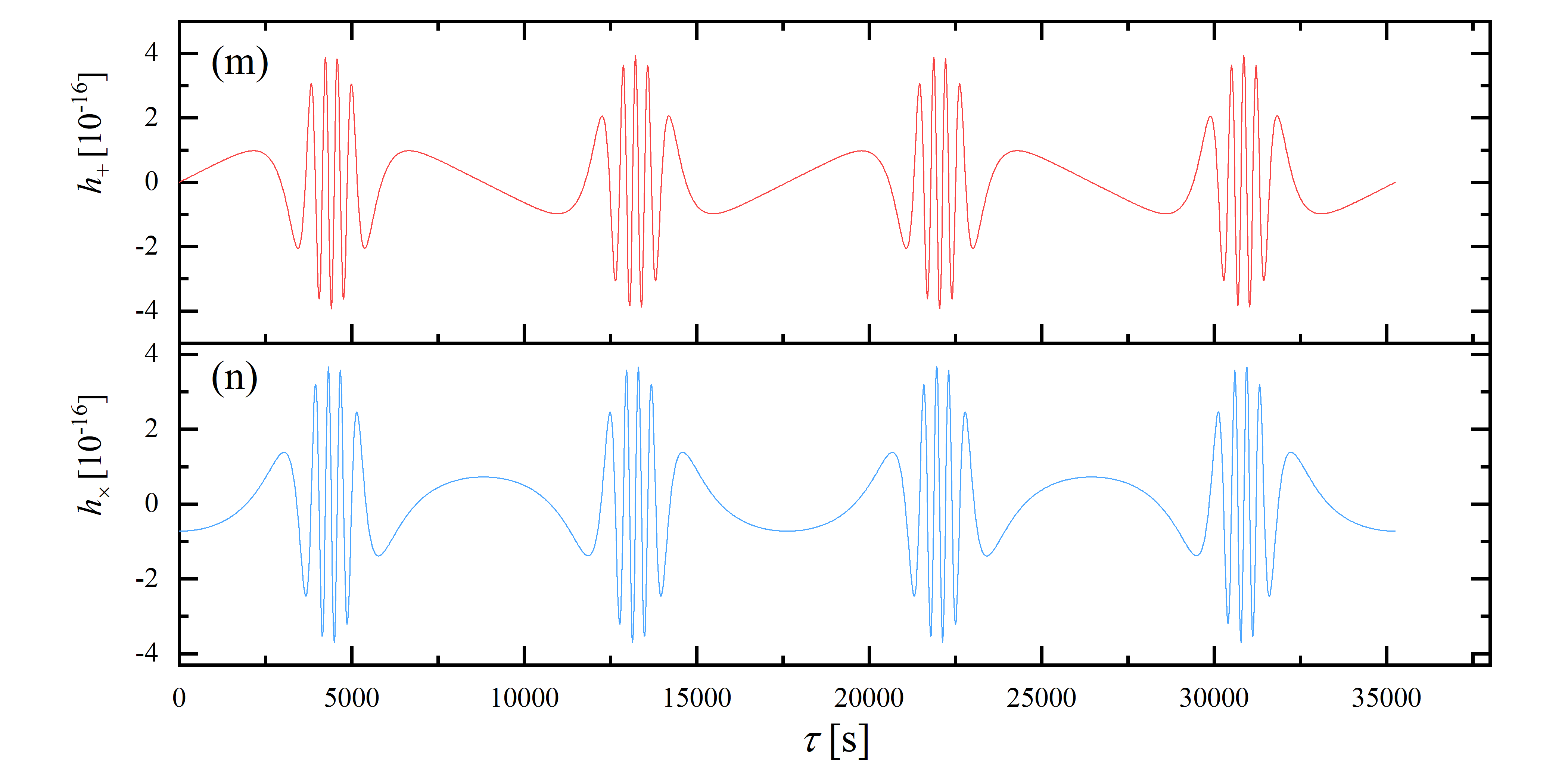}
\includegraphics[width=7.5cm]{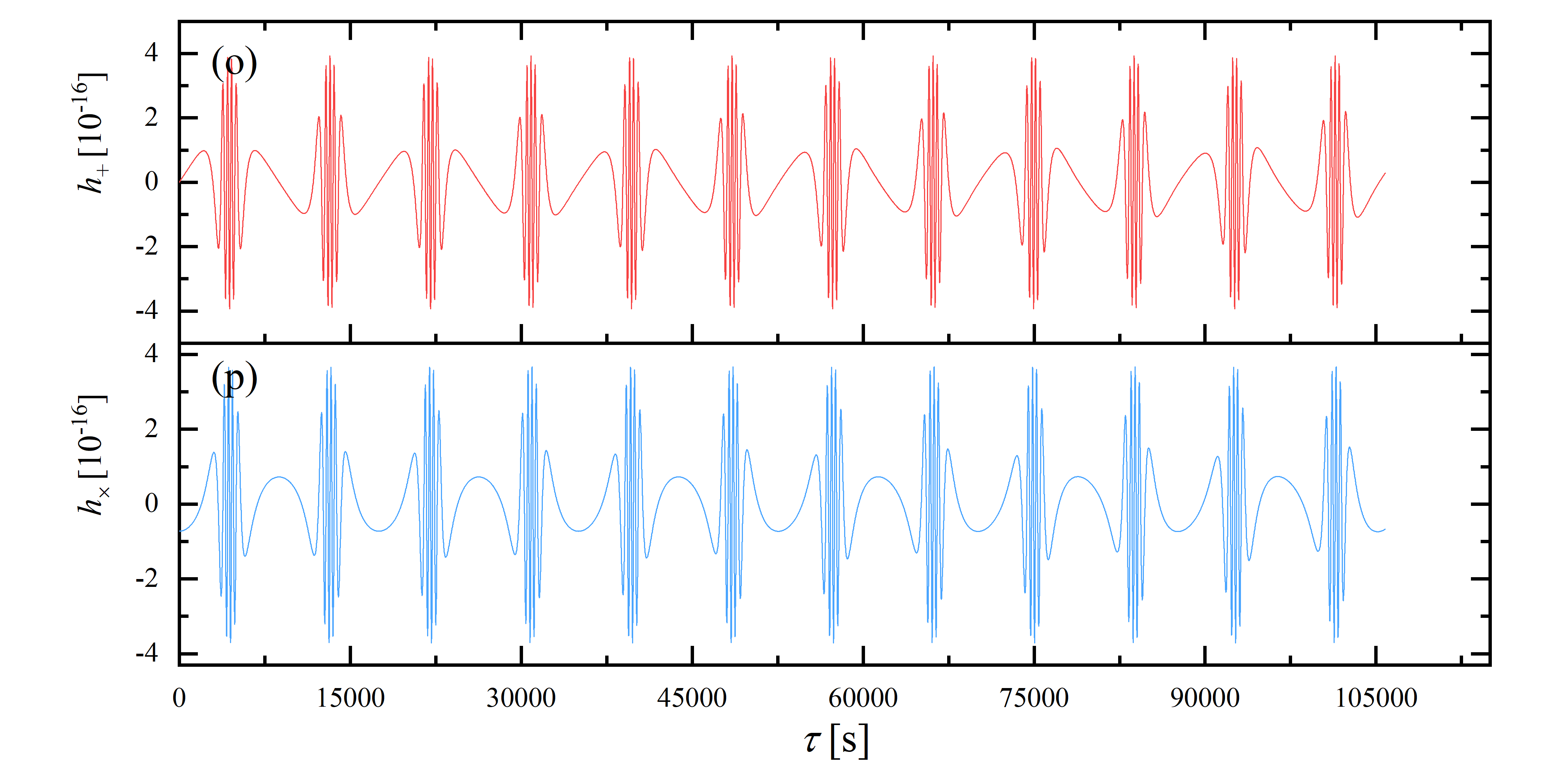}
\includegraphics[width=7.5cm]{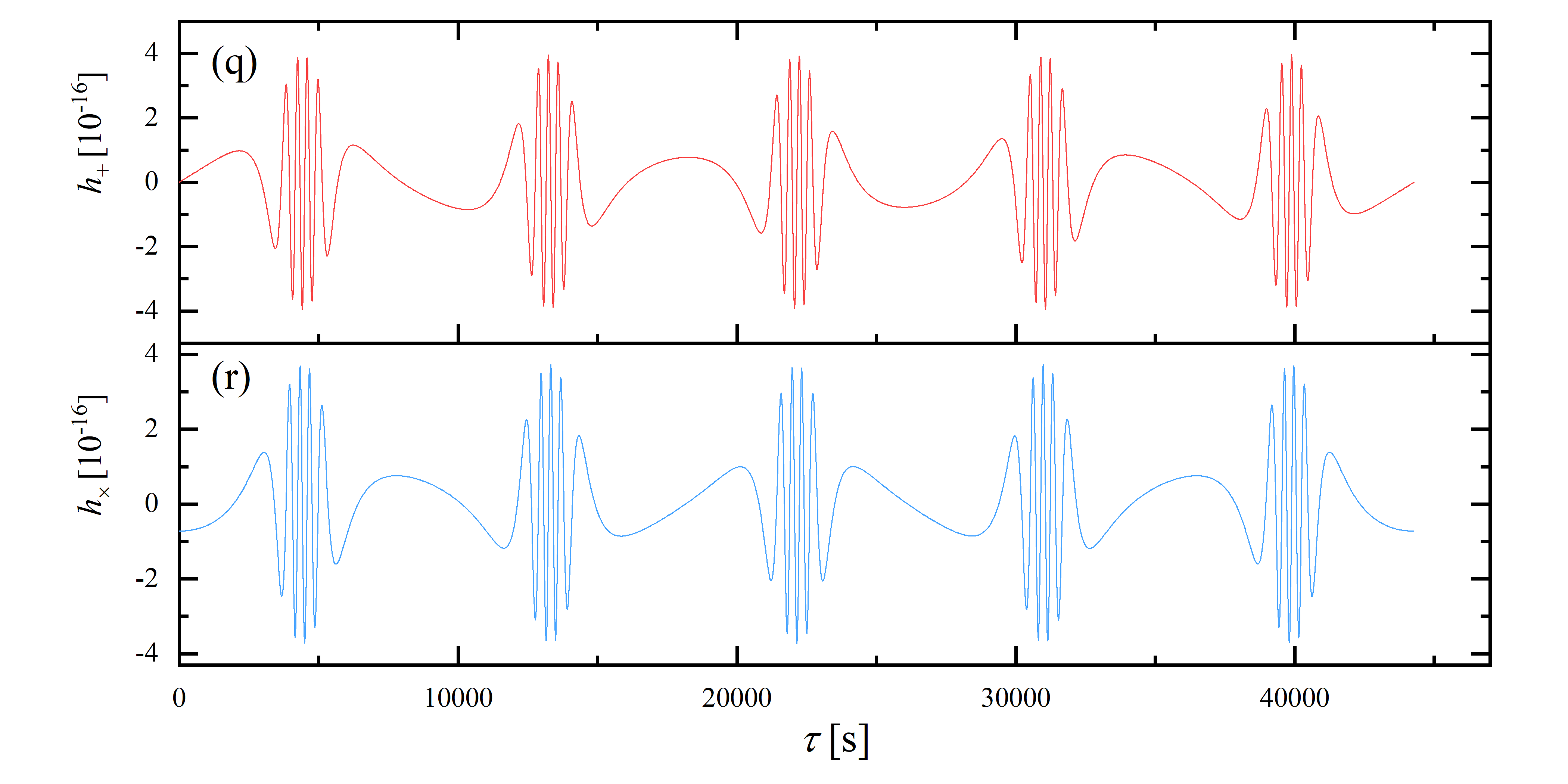}
\includegraphics[width=7.5cm]{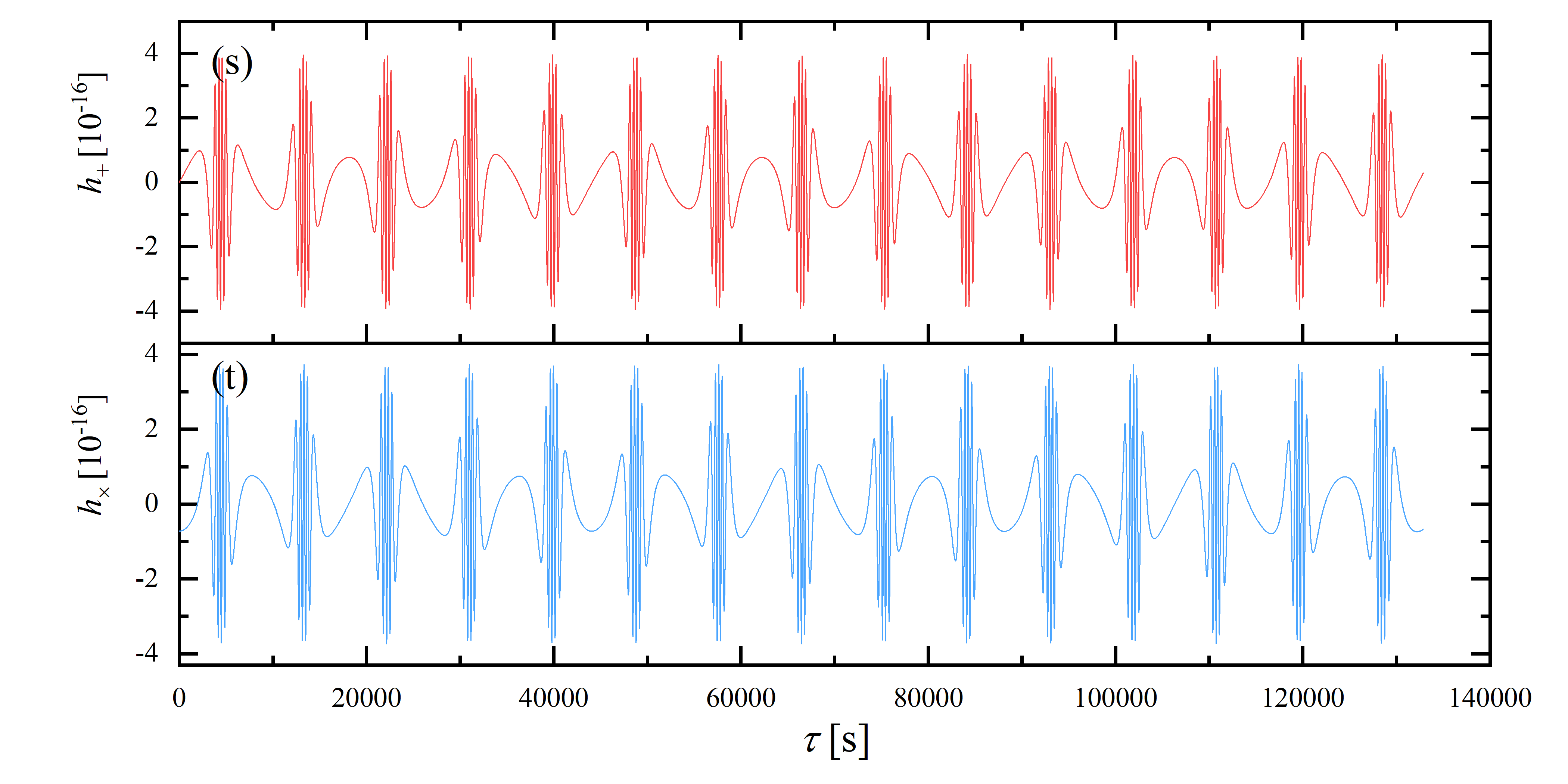}
\caption{Time-domain gravitational wave signals of rational orbits (left column) and irrational orbits (right column). From top to bottom $z$ sequentially increases from $1$ to $5$. In each panel, red and blue correspond to the $h_{+}$ and $h_{\times}$ polarization states respectively. From the left column, it can be concluded that within one orbital period, the number of high-frequency oscillations in the gravitational wave signal is consistent with the number of orbital leaves.}      
\label{fig6}                 
\end{figure*}

Upon obtaining the orbits, we fix $\iota$ and $\zeta$ to $\pi/4$. By utilizing equations \eqref{19} and \eqref{20}, the code returns the two gravitational wave polarization states $h_{+}$ and $h_{\times}$ corresponding to the orbits. Meanwhile, since the gravitational waves possess observational potential, in this step we restore the geometrized units to the International System of Units, which requires specifying the concrete parameters of the source. We adopt the background of the supermassive black hole at the Galactic Center, with a black hole mass $M \approx 4 \times 10^{6}$ $M_{\odot}$ and a distance $D_{\textrm{L}} \approx 8000$ pc, and assume the mass of the timelike small celestial body to be $m = 100$ $M_{\odot}$. Then, the intensity scaling factor of the gravitational waves is $\eta GM/(D_{\textrm{L}}c^{2}) \approx 5.9887 \times 10^{-16}$, and the time scaling factor is $GM/c^{3} \approx 19.7088$. Figure \ref{fig6} presents the gravitational wave polarization states corresponding to the orbits of various configurations in figure \ref{fig5}, where the simulation time for rational orbits (left column) is exactly one orbital period, and that for irrational orbits is three times the period of the rational orbits. It is not difficult to find that the oscillatory behaviors of the two gravitational wave polarization states are similar, and the evolution of their amplitudes is also close. The gravitational wave waveforms are all composed of two components: one is the high-frequency oscillatory component, such as the segment where $\tau=4000 \sim 5000$ in panel (a); the other is the relatively flat region containing broad peaks that connects the high-frequency oscillations, such as the segment where $\tau=5000 \sim 12000$ in panel (c). Some of the authors of this paper pointed out in \cite{Tan:2026yjg} that the high-frequency oscillatory component originates from the orbital motion near the periastron, while the flat region arises from the contribution of the motion of the small celestial body near the apastron. Therefore, we can observe that the number of high-frequency oscillations in the waveform perfectly matches the number of orbital leaves $z$. On the other hand, we find that the gravitational wave waveforms of rational orbits and irrational orbits are almost identical within a relatively short period of time.

\begin{figure*}
\centering                   
\includegraphics[width=3cm]{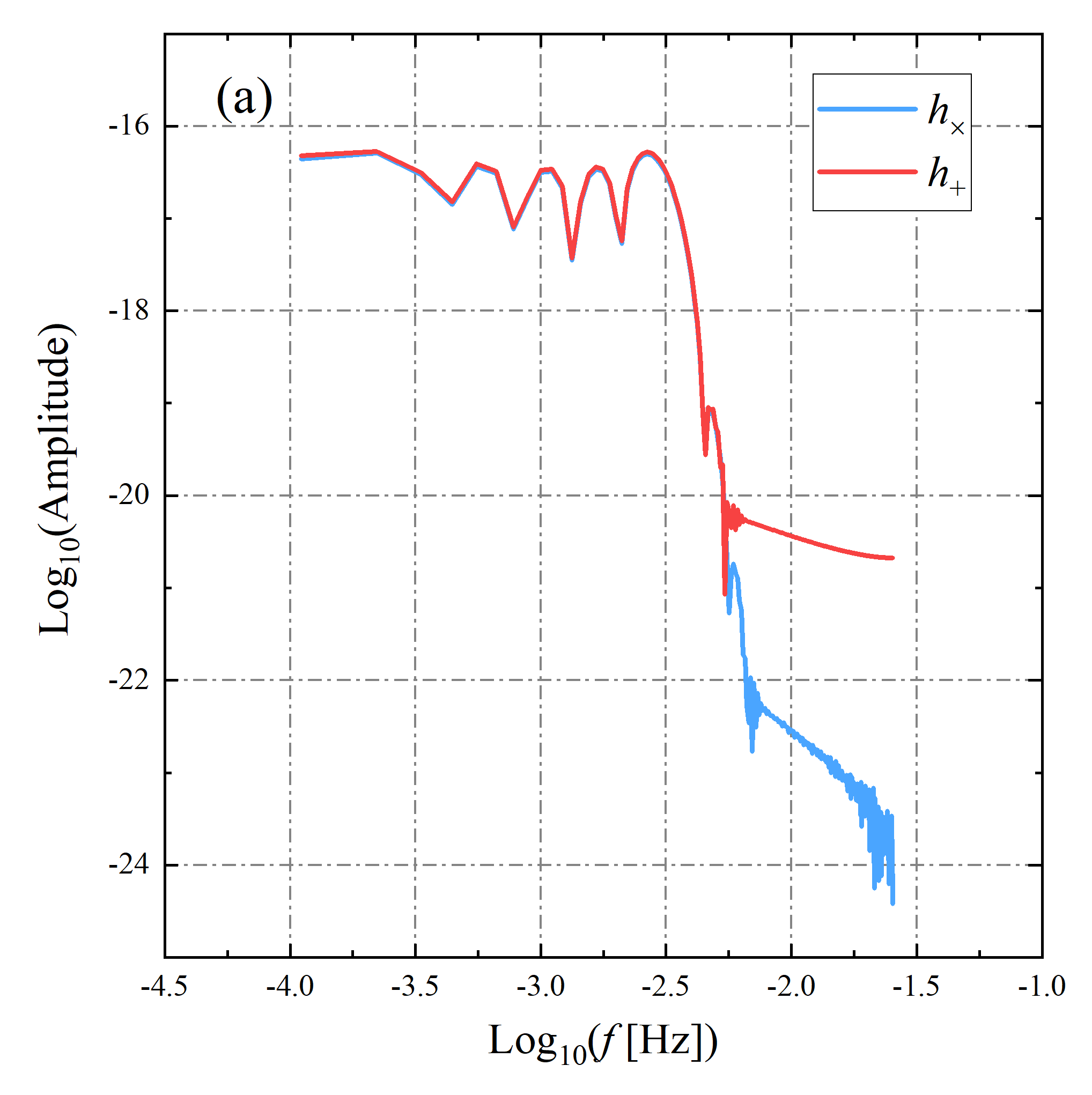}
\includegraphics[width=3cm]{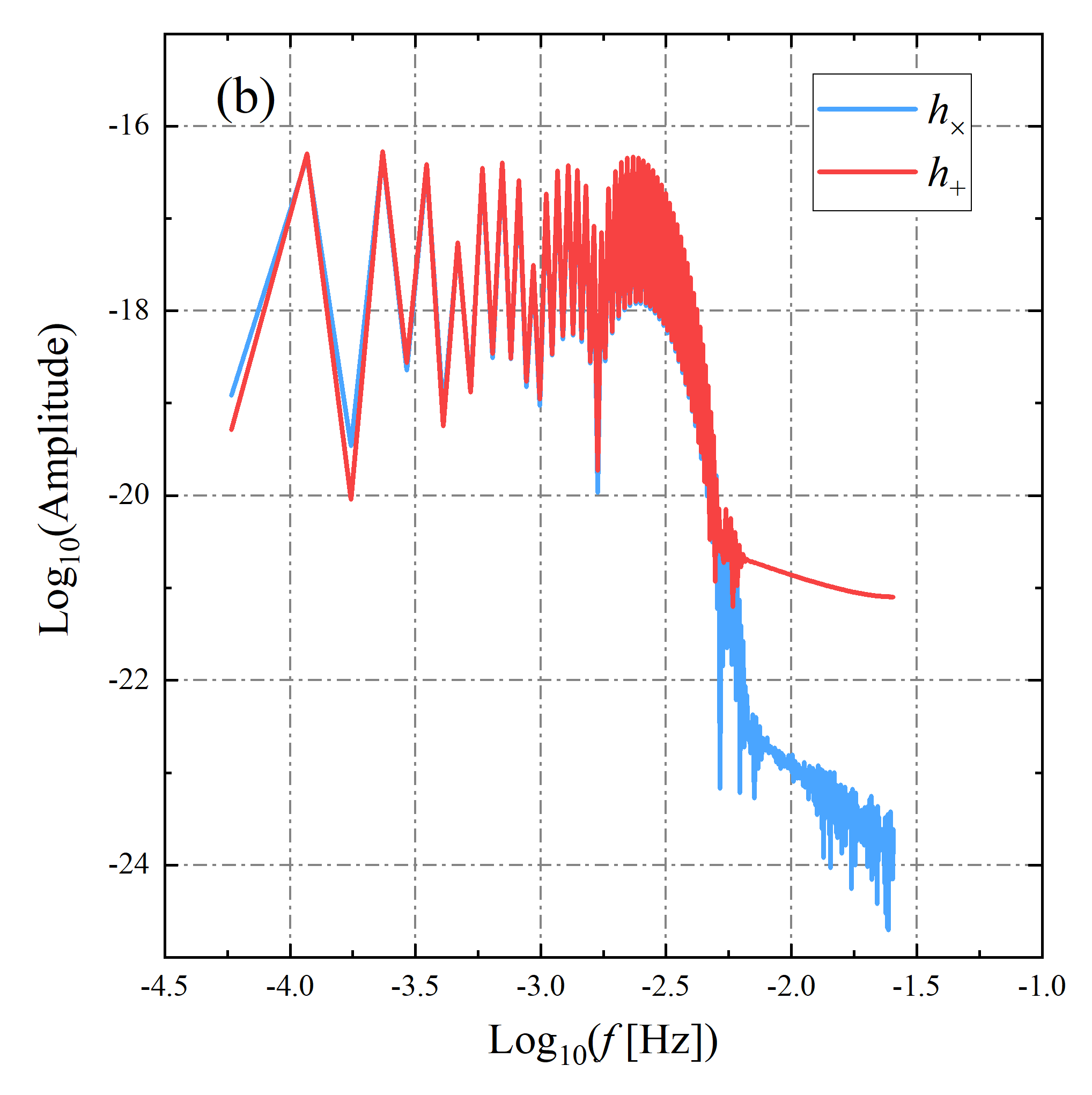}
\includegraphics[width=3cm]{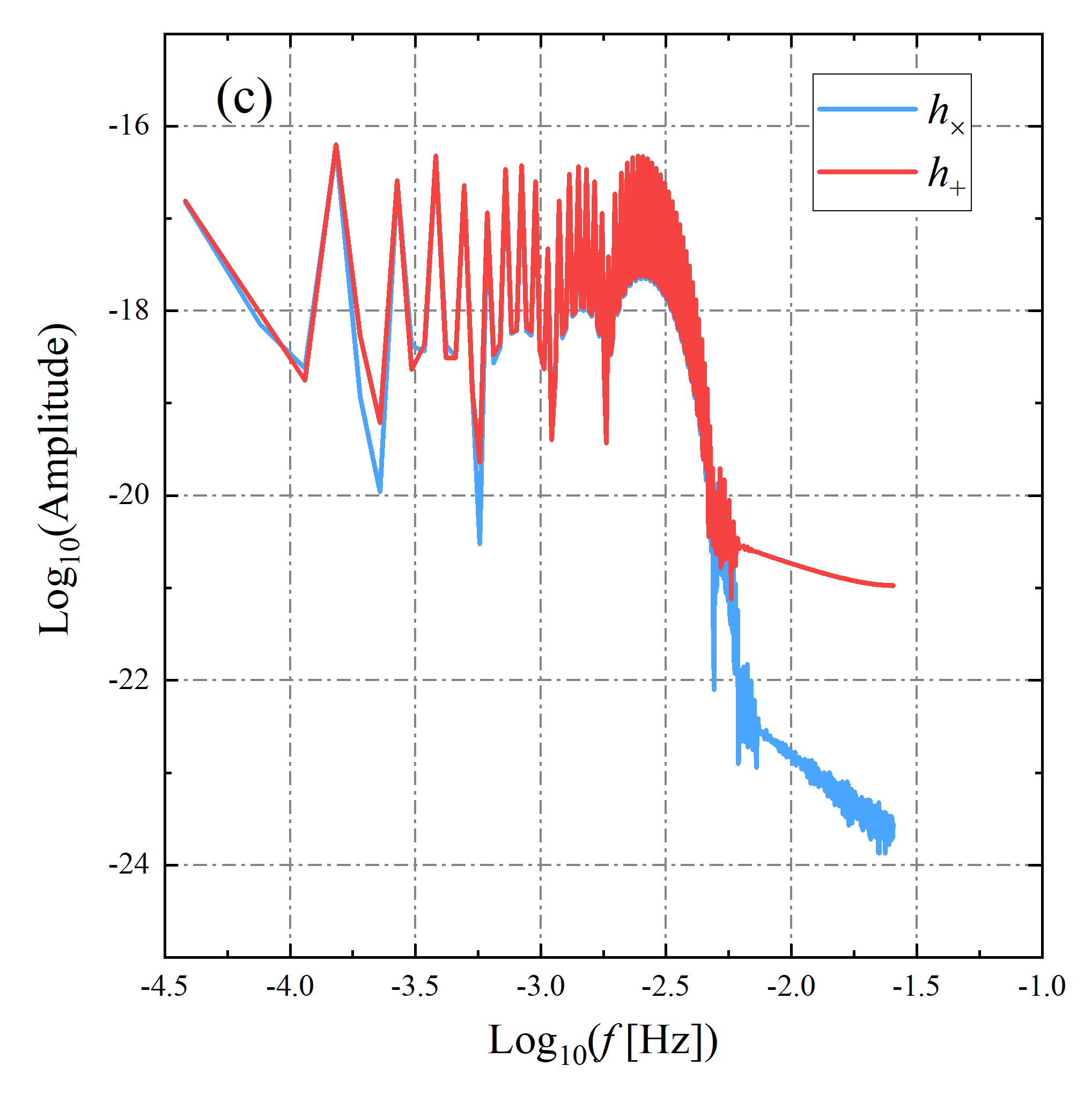}
\includegraphics[width=3cm]{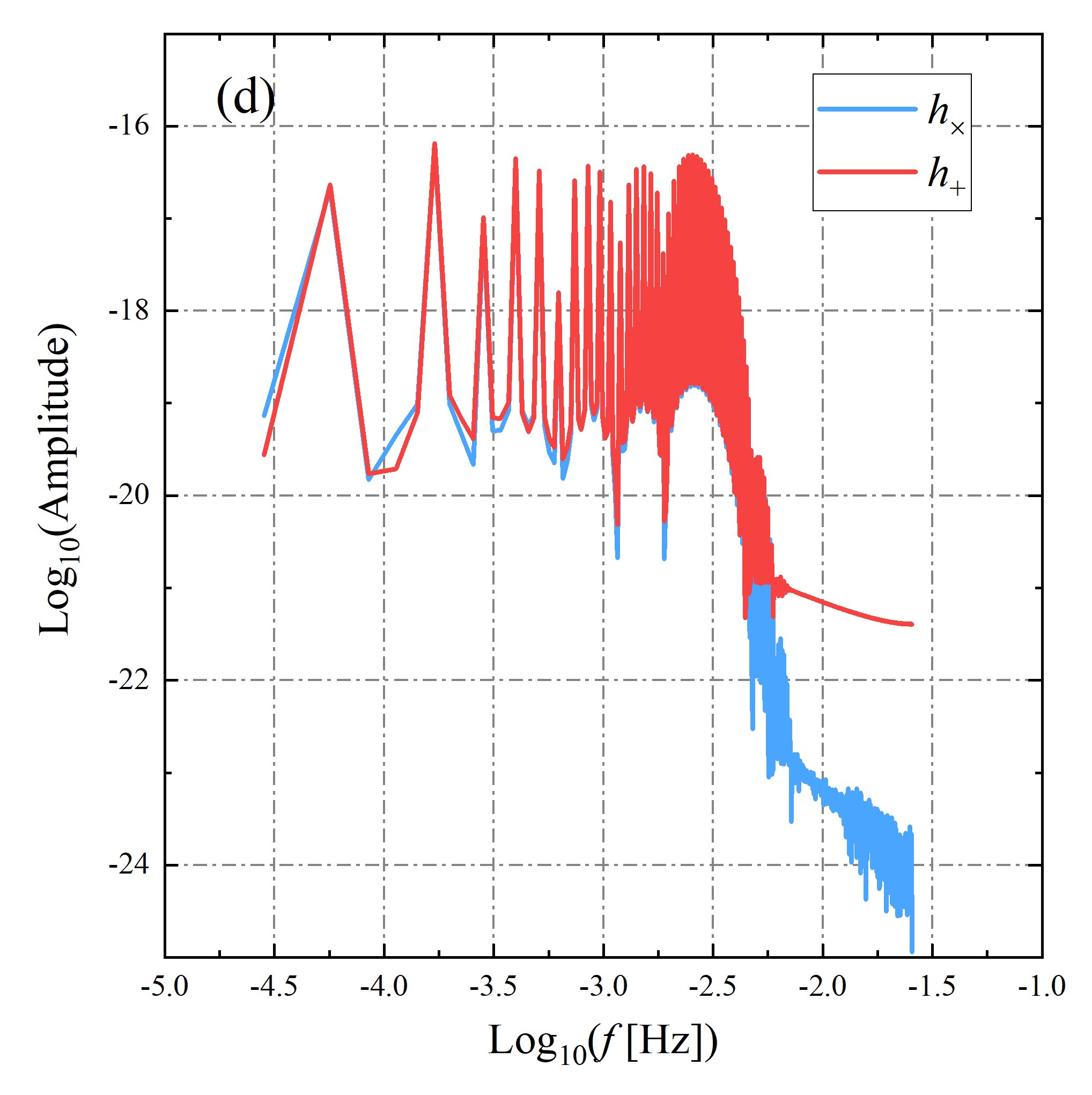}
\includegraphics[width=3cm]{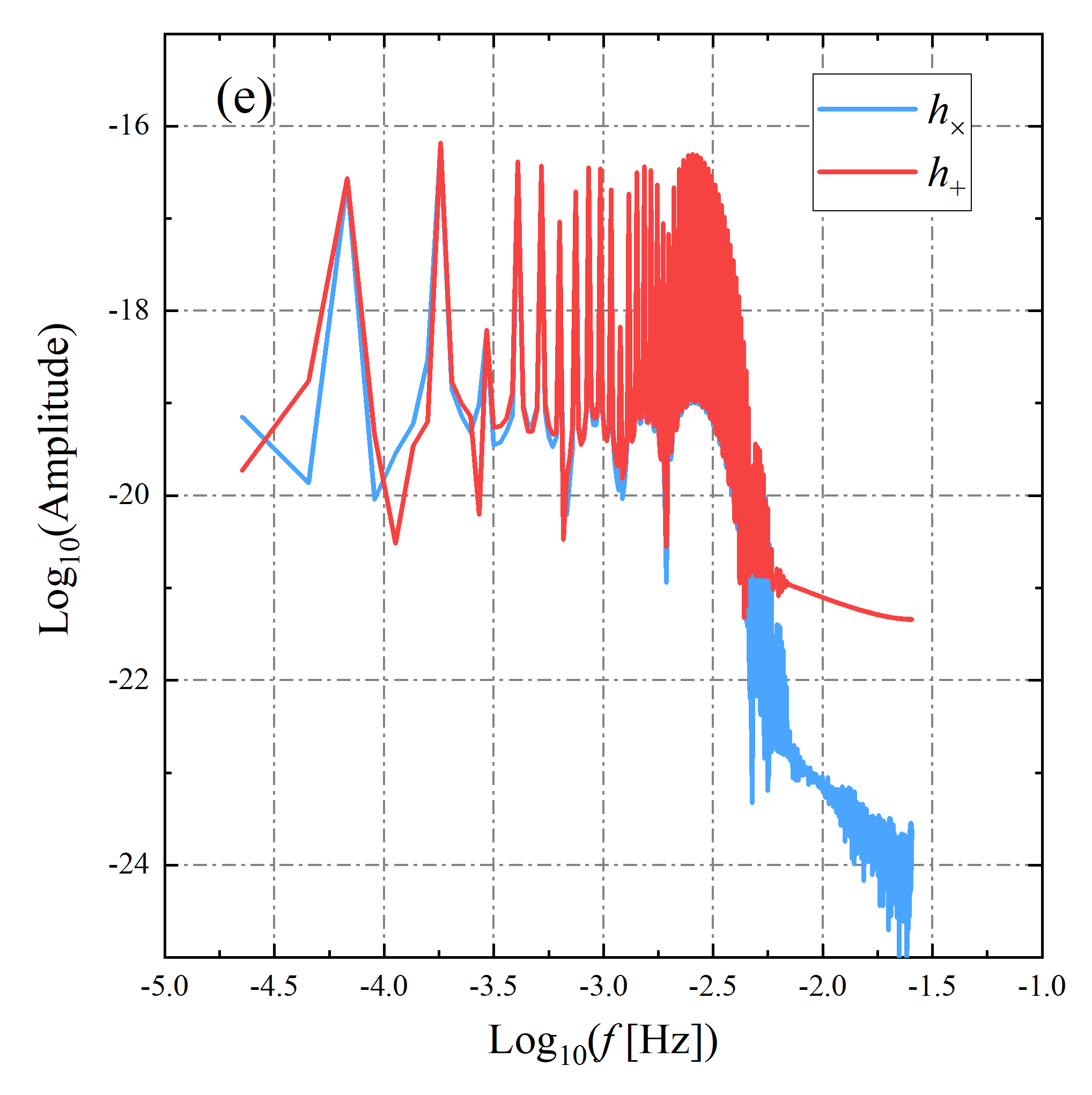}
\caption{Frequency-domain information obtained via the fast Fourier transform from the time-domain gravitational wave signals of the rational orbits in figure \ref{fig6}. From left to right, the number of orbital leaves $z$ sequentially increases from $1$ to $5$. It can be observed that the frequency-domain amplitudes of the gravitational waves are relatively strong below $0.01$ Hz. Meanwhile, it is almost impossible to establish a connection between the frequency-domain gravitational wave signals and the orbital configurations.}      
\label{fig7}                 
\end{figure*}

Our code employs the fast Fourier transform to process the time-domain signals of the gravitational waves to obtain information in the frequency-domain. Figure \ref{fig7} displays the frequency-domain signals of the two gravitational wave polarization states for the rational orbits in the first row of figure \ref{fig5}, where red corresponds to the ``plus'' polarization state and blue corresponds to the ``cross'' polarization state. It is found that the signal windows of both polarization states range from $10^{-4} \sim 10^{-1.5}$ Hz, within which the signal strength at the mHz level and below remains robust. Simultaneously, we discover that the amplitude range of $h_{\times}$ is larger than that of $h_{+}$, and the former can even drop as low as $10^{-25}$ when the frequency $f$ is higher than the order of $10^{-2}$ Hz. Furthermore, it is necessary to point out that it is rather difficult to identify the orbital characteristics using solely the frequency-domain signals of the gravitational waves.

\begin{figure*}
\centering                   
\includegraphics[width=7cm]{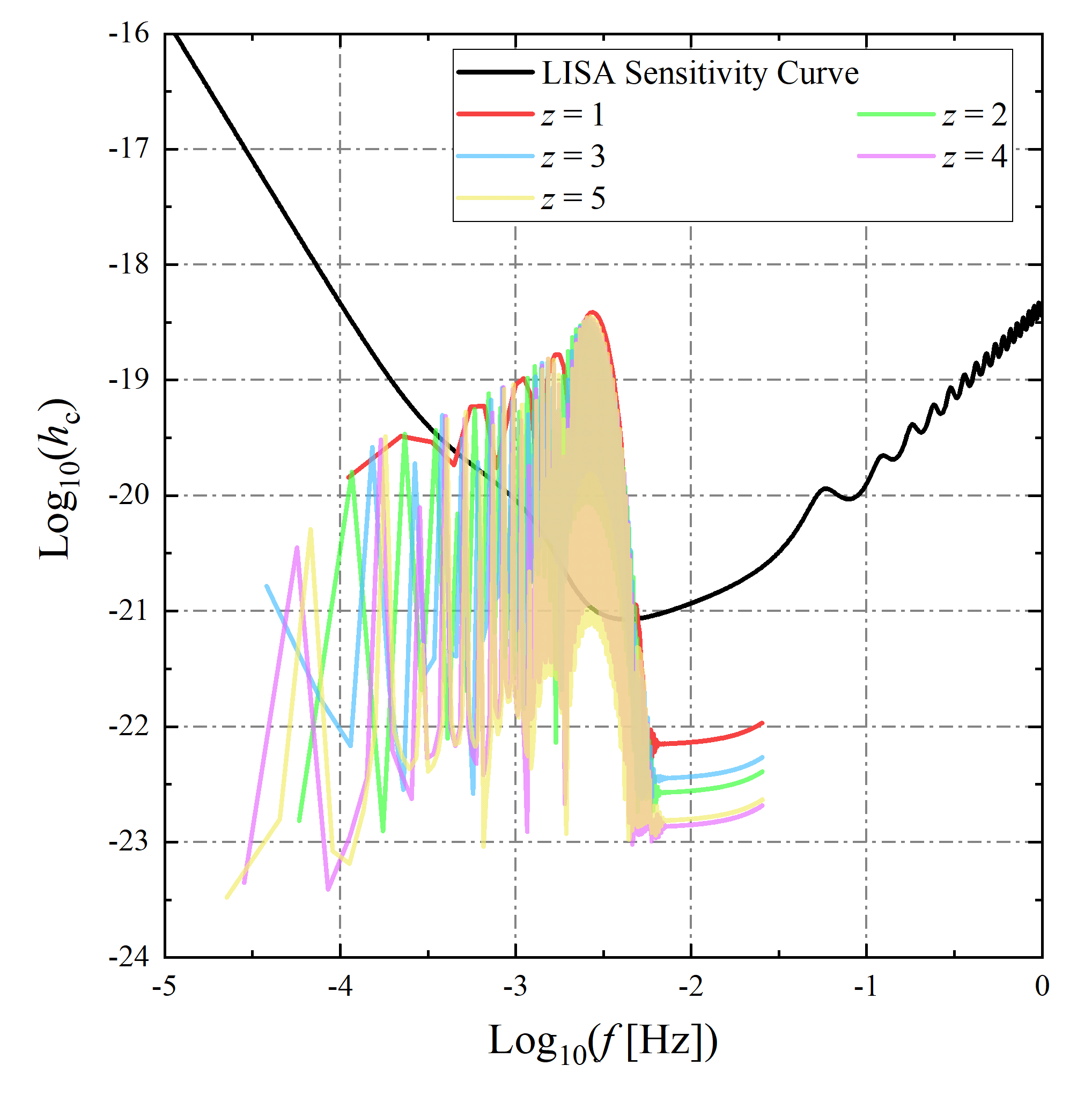}
\caption{Comparison between the characteristic strains of gravitational waves for different orbital configurations and the LISA sensitivity curve (black). It can be found that the characteristic strains of all the displayed orbital configurations have portions exceeding the LISA sensitivity curve, and these correspond to the mHz frequency window. This demonstrates that the gravitational radiation of the rational orbits investigated in this paper has the potential to be observed in the future.}      
\label{fig8}                 
\end{figure*}

Utilizing equation \eqref{21}, our numerical tool further obtains the characteristic strain $h_{\textrm{c}}$ of the gravitational waves for each orbit, as shown in figure \ref{fig8}, where the black curve represents the sensitivity curve of LISA \cite{Robson:2018ifk}. It can be observed that, although the segments are relatively small, the characteristic strain curves of the gravitational waves for the five rational orbits exhibit portions that are higher than the LISA sensitivity curve. This indicates that the gravitational radiation from rational orbits possesses potential observational viability. On the other hand, the background of our model is a dynamical system composed of an intermediate mass black hole and the Galactic Center supermassive black hole. The fact that the gravitational wave signals of this system can be captured by LISA implies that there is a theoretical foundation for utilizing LISA to search for intermediate mass black holes near the Galactic Center.
\section{Conclusion and Discussion}
In this paper, based on Mathematica and the OpenMP framework, we have developed a numerical code for solving rational orbits and irrational orbits applicable to any static spherically symmetric spacetime, which also supports the further calculation of gravitational wave signals in the time-domain and frequency-domain, as well as the characteristic strains\footnote{Our code is now publicly available on GitHub at https://github.com/Shiyang-Hu/Rational-Orbits-and-Gravitational-Radiation-in-Spherically-Symmetric-Spacetimes.}. Users only need to define the covariant metric of the target spacetime and set the parameters of the source to operate it smoothly. Meanwhile, our code exhibits high execution efficiency. On a platform with an Intel i7--14700 processor and 8 GB RAM, simulating and plotting the data from figure \ref{fig3} $\sim$ figure \ref{fig8} takes only $65$ seconds. This numerical tool possesses considerable guiding value in the fields of orbital dynamics in curved spacetimes and gravitational wave astronomy.

Utilizing this code, we have demonstrated the rational orbits and irrational orbits of various orbital configurations in the Schwarzschild spacetime, and discovered that the number of high-frequency oscillations in the time-domain gravitational wave waveforms perfectly matches the number of orbital leaves. More importantly, by comparing the characteristic strains of the gravitational radiation from each orbit with the LISA sensitivity curve, we find that the gravitational waves radiated by extreme-mass-ratio inspiral systems composed of an intermediate mass black hole and the Galactic Center supermassive black hole are promising targets for future observation. This further demonstrates that utilizing LISA to search for intermediate mass black holes near the Galactic Center is feasible.

Furthermore, it is worth mentioning that astrophysical black holes often inherit the angular momentum of their progenitor stars and consequently possess spin. Their spacetimes are therefore predominantly characterized by axisymmetric geometries. Thus, it is highly necessary to extend the applicable scope of our code to rotating spacetimes. In this procedure, the crucial step is to adapt the calculations for the timelike particle radial equation \eqref{9}, the effective potential \eqref{10}, and the azimuthal angle accumulation $\Delta\varphi$ to their counterparts for rotating black holes. In future work, we plan to expand the capability of this code to automatically search for rational orbits within rotating spacetimes.
\section*{Acknowledgments}
The authors are very grateful to the referee for insightful comments and valuable suggestions. This work was supported by the National Natural Science Foundation of China under Grant No. 12403081, the Natural Science Foundation of Hunan Province under Grant No. 2026JJ50351, and the Scientific Research Foundation of Hunan Provincial Education Department under Grant No. 25B0373.

\bibliographystyle{iopart-num}
\bibliography{references}
\end{document}